\documentclass[aps,prd,twocolumn,floatfix,superscriptaddress,showpacs,amsmath]{revtex4-1}
\usepackage{graphicx}
\usepackage{bm}
\usepackage{color}

\begin{document}

\title{Effects of neutrino oscillations on nucleosynthesis and neutrino signals for 
an $18\,M_\odot$ supernova model}

\author{Meng-Ru Wu}
\affiliation{Institut f{\"u}r Kernphysik (Theoriezentrum), Technische
  Universit{\"a}t Darmstadt, Schlossgartenstra{\ss}e 2, 64289
  Darmstadt, Germany} 
\affiliation{School of Physics and Astronomy, University of Minnesota,
  Minneapolis, MN 55455, USA}  
 \author{Yong-Zhong Qian}
\affiliation{School of Physics and Astronomy, University of Minnesota,
  Minneapolis, MN 55455, USA}
\author{Gabriel Mart{\'i}nez-Pinedo}
\affiliation{Institut f{\"u}r Kernphysik (Theoriezentrum), Technische
  Universit{\"a}t Darmstadt, Schlossgartenstra{\ss}e 2, 64289
  Darmstadt, Germany} 
\affiliation{GSI Helmholtzzentrum f\"ur Schwerioneneforschung,
  Planckstra{\ss}e~1, 64291 Darmstadt, Germany} 
\author{Tobias Fischer}
\affiliation{Institute for Theoretical Physics, University of Wroc{\l}aw,
pl. M. Borna 9, 50-204 Wroc{\l}aw, Poland} 
\author{Lutz Huther}
\affiliation{Institut f{\"u}r Kernphysik (Theoriezentrum), Technische
  Universit{\"a}t Darmstadt, Schlossgartenstra{\ss}e 2, 64289
  Darmstadt, Germany} 

\date{\today}

\begin{abstract}
In this paper, we explore the effects of neutrino flavor oscillations
on supernova nucleosynthesis and on the neutrino signals. Our study is
based on detailed information about the neutrino spectra and their
time evolution from a spherically-symmetric supernova model for an
$18\,M_\odot$ progenitor. We find that collective neutrino oscillations are not
only sensitive to the detailed neutrino energy and angular
distributions at emission, but also to the time evolution of both the
neutrino spectra and the electron density profile. We apply the results of
neutrino oscillations to study the impact on supernova
nucleosynthesis and on the neutrino signals from a Galactic supernova. We
show that in our supernova model, collective neutrino oscillations
enhance the production of rare isotopes $^{138}$La and $^{180}$Ta but have
little impact on the $\nu$p-process nucleosynthesis. In addition, the adiabatic MSW
flavor transformation, which occurs in the C/O and He shells of the
supernova, may affect the production of light nuclei such as $^7$Li and
$^{11}$B. For the neutrino signals, we calculate the rate of neutrino
events in the Super-Kamiokande detector and in a hypothetical liquid
argon detector. Our results suggest the possibility of using
the time profiles of the events in both detectors, along with the
spectral information of the detected neutrinos, to infer the neutrino mass
hierarchy.
\end{abstract}


\maketitle

\section{Introduction}
Core-collapse supernovae signify the death of massive stars heavier than 
$\sim 8\,M_\odot$ and the birth of proto-neutron stars. In each explosion 
$\sim 10^{53}$~erg of gravitational binding energy is released through 
emission of $\sim 10^{58}$ neutrinos (antineutrinos) of all three flavors
over $\sim 10$~s. These neutrinos play essential roles in the dynamics and 
nucleosynthesis of supernovae. Prominent examples include 
revival of the stalled supernova shock by neutrino heating in conjunction with 
fluid instabilities (\cite{Bethe:1984ux}; see \cite{Janka:2012wk} for a review), 
production of heavy elements in neutrino-driven winds from proto-neutron stars
(e.g., \cite{Qian:1996xt}; see \cite{Arcones:2012wj} for a recent review), 
and neutrino-induced nucleosynthesis in outer shells of supernovae
(e.g., \cite{1988PhRvL..61.2038E,1990ApJ...356..272W,Banerjee:2011zm,2013PhRvL.110n1101B}). 
In addition, current and planned 10-kiloton-scale detectors are able to observe 
thousands of neutrino events if a supernova occurs in the Galaxy 
(see \cite{Scholberg:2012id} for a review). Such detection would provide 
a unique opportunity to explore the physics of core-collapse supernovae 
and properties of neutrinos.

In the absence of flavor oscillations, we would only need the emission 
characteristics of neutrinos determined by their decoupling from the 
proto-neutron star, such as their luminosities, energy spectra, and 
angular distributions, in order to understand their roles in supernovae. It would also 
be straightforward to infer the neutrino luminosities and energy spectra at emission
from signals in appropriate detectors for a Galactic supernova. 
However, neutrino oscillations have been established by various experiments.
Consequently, we must take neutrino flavor evolution into account when assessing 
the effects of neutrinos on the dynamics and nucleosynthesis of supernovae and
when deciphering the rich underlying physics from supernova neutrino signals.
In this paper we present a framework for calculating neutrino flavor 
evolution in the dynamic supernova environment, perform detailed calculations
for an $18\,M_\odot$ supernova model, and examine the effects of neutrino
oscillations on nucleosynthesis and neutrino signals for this model.

The intrinsic parameters describing neutrino oscillations include three vacuum
mixing angles ($\theta_{12}$, $\theta_{13}$, $\theta_{23}$), a $CP$-violating 
phase ($\delta_{CP}$), and two independent mass-squared-differences (e.g., 
$\Delta m_{21}^2\equiv m_2^2-m_1^2$, $\Delta m_{31}^2\equiv m_3^2-m_1^2$) 
between neutrino vacuum mass eigenstates. Observations of solar and 
atmospheric neutrinos and other terrestrial experiments have measured 
$\theta_{12}$, $\theta_{13}$, $\theta_{23}$, $\Delta m_{21}^2$, and
$|\Delta m_{31}^2|$ to good precision (see review in \cite{Agashe:2014kda}). 
There are ongoing and planned experiments 
to measure the yet unknown $\delta_{CP}$ and sign of 
$\Delta m_{31}^2$. The latter is also referred to as the neutrino mass hierarchy,
with $\Delta m_{31}^2>0\ (< 0)$ defined as normal (inverted). In this paper we 
assume $\delta_{CP}=0$ but consider both normal and inverted mass hierarchies.

\begin{figure}
\centering
\includegraphics[width=0.8\columnwidth]{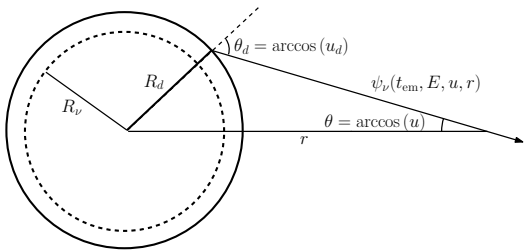}
\caption{Sketch of neutrino emission and propagation.}
\label{Fig-sketch}
\end{figure}

We divide the supernova environment into two regions separated by a
``decoupling sphere'' at radius $r=R_d$. We assume that at $r<R_d$,
neutrino interactions with matter dominate and flavor oscillations 
have no net effect. Classical Boltzmann transport equations coupled with 
supernova conditions then determine the energy and angular distributions 
$f_\nu(t_{\rm em},E,\theta_d,R_d)$ for neutrinos emitted at $r=R_d$, where 
$t_{\rm em}$ is the time of emission, $E$ is the neutrino energy, and $\theta_d$ 
is the angle of propagation with respect to the radial direction at $r=R_d$ 
(see Fig.~\ref{Fig-sketch}). An important feature of these distributions is
the hierarchy of the corresponding average neutrino energies
$\langle E_{\nu_e}\rangle <\langle E_{\bar\nu_e}\rangle <
\langle E_{\nu_{\mu(\tau)}}\rangle\approx\langle E_{\bar\nu_{\mu(\tau)}}\rangle$.
At $r>R_d$, only a small fraction 
of neutrinos can still interact with matter to affect supernova dynamics and 
nucleosynthesis. For the purpose of treating neutrino flavor evolution, we assume 
that all neutrinos are free-streaming at $r>R_d$.

Flavor evolution of neutrinos at $r>R_d$ can exhibit rich phenomena because
they propagate through an enormous range of matter density, the structure of
which may be complicated by convection-driven fluctuations and propagation
of the supernova shock. These factors influence neutrino oscillations through the
Mikheyev-Smirnov-Wolfenstein (MSW) effect induced by forward scattering of 
neutrinos on electrons \cite{Wolfenstein:1977ue,Mikheyev:1985}. In addition, 
due to the non-linear coupling through forward scattering of neutrinos on other 
neutrinos, collective oscillations among all three flavors of neutrinos 
(antineutrinos) may occur within $\sim 100$~km of the proto-neutron star 
(see \cite{Duan:2010bg} for a review and 
\cite{Raffelt:2011yb,Duan:2010bf,Chakraborty:2011nf,Chakraborty:2011gd,
Galais:2011gh,Wu:2011yi,Mirizzi:2011tu,Pehlivan:2011hp,Banerjee:2011fj,
Sarikas:2011am,Sarikas:2012vb,Saviano:2012yh,Cherry:2012zw,Volpe:2013uxl,
Raffelt:2013rqa,Raffelt:2013isa,Cherry:2013mv,Vlasenko:2013fja,Mirizzi:2013rla,
Duan:2013kba,Chakraborty:2014nma,Akhmedov:2014ssa,Vlasenko:2014bva} 
for more recent developments). In any case, as $\nu_\mu$, $\bar\nu_\mu$, 
$\nu_\tau$, and $\bar\nu_\tau$ have higher average energies than $\nu_e$ and 
$\bar\nu_e$ at emission, flavor oscillations at $r>R_d$ may have important 
effects on supernova dynamics, nucleosynthesis, and neutrino signals. 

While neutrino oscillations in supernovae have been studied extensively,
our approach in this paper differs from these previous works in that we
employ neutrino emission characteristics and electron number density 
profiles calculated self-consistently by a supernova model and that we
explicitly take the time evolution of these quantities into account 
when calculating neutrino flavor evolution through the supernova 
environment over the period of significant neutrino emission.
The following example illustrates why such an approach is required to 
adequately examine the impact of neutrino oscillations on supernova physics. 
Consider a mass element moving along a radial trajectory $r_m(t)$, and 
for simplicity, ignore the neutrino travel time from emission to reaching the 
mass element. To calculate the rates of neutrino reactions in this mass 
element, we need quantities such as $P_{\nu_e\nu_e}(t_{\rm em},E,\theta,r)$
for $t_{\rm em}=t$ and $r=r_m(t)$, which gives the survival probability for a 
$\nu_e$ emitted with energy $E$ at time $t_{\rm em}$ and arriving
at radius $r>R_d$ with an angle of propagation $\theta$ with respect to the 
radial direction (see Fig.~\ref{Fig-sketch}). 
As the process of nucleosynthesis in the mass element can last up to
$\sim 10$~s and there are large changes in $r_m$, neutrino emission 
characteristics, and the electron number density profile over this time,
we must calculate $P_{\nu_e\nu_e}(t_{\rm em},E,\theta,r)$ and similar 
survival probabilities for many time snapshots of the supernova input for 
neutrino flavor evolution. Therefore, our results on neutrino oscillations 
are given in terms of these survival probabilities on an extensive 
four-dimensional grid covering wide ranges of emission time,
neutrino energy, propagation angle, and arrival radius. 
Our methodology is demonstrated for a specific $18\,M_\odot$ supernova 
model, and can be generalized to any spherically-symmetric models.

We find that collective oscillations are sensitive to the details of the 
neutrino energy and angular distributions at emission and to the
time evolution of these distributions and the electron number 
density profile. For the specific model studied, although collective 
neutrino oscillations occur too far out to affect nucleosynthesis in 
the neutrino-driven wind, they can still affect neutrino-induced
nucleosynthesis in outer supernova shells in combination with the
MSW effect. We show that for a Galactic supernova described by
the same model, the neutrino signals are mainly modified by the
MSW effect and those signals during shock revival can be used to 
infer the yet unknown neutrino mass hierarchy. 

We describe the supernova model in
Sec.~\ref{sec-model} and our approach to calculate neutrino 
flavor evolution in Sec.~\ref{sec-method}. We present and discuss
our results on neutrino oscillations in Sec.~\ref{sec-nuosc}.
We apply these results to assess the effects of neutrino
oscillations on nucleosynthesis in Sec.~\ref{sec-nucleo}
and to analyze the neutrino signals in Super-Kamiokande and
a hypothetical liquid argon detector in Sec.~\ref{sec-signal}.
We discuss all our results and conclude in Sec.~\ref{sec-summary}.

\section{Supernova Model}\label{sec-model}
\begin{figure}
\centering
\includegraphics[width=0.9\columnwidth, angle=-90]{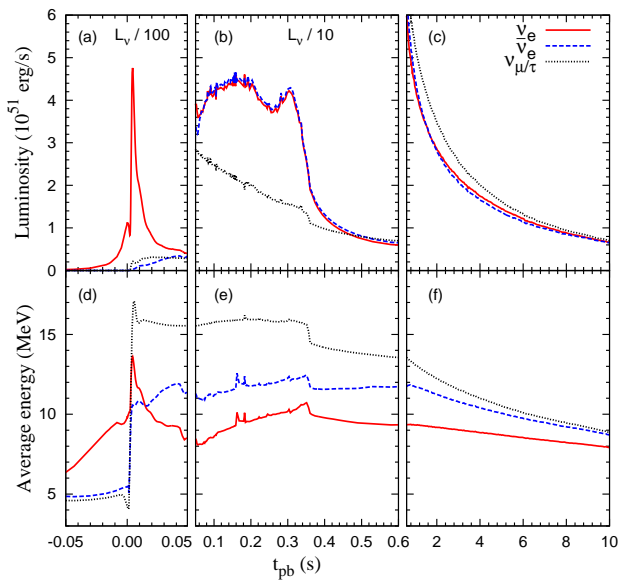}
\caption{Evolution of neutrino luminosities [panels (a)--(c)]
and average energies [panels (d)--(f)] for the $18\,M_\odot$ 
supernova model as observed at infinity \cite{Fischer:2009af}.}
\label{Fig-lumin}
\end{figure}

We adopt a supernova model with an $18\,M_\odot$ progenitor.
This model is based on general-relativistic radiation 
hydrodynamics in spherical symmetry and incorporates
detailed three-flavor Boltzmann neutrino transport \cite{Fischer:2009af}.
The core collapse is initiated by loss of energy and pressure through 
photo-disintegration of iron-group nuclei and capture of electrons on 
protons and nuclei. Neutrinos produced during the collapse are 
predominantly $\nu_e$, which are trapped and can only diffuse 
out of the core with $\langle E_{\nu_e}\rangle\approx 4$--9~MeV
(see Fig.~\ref{Fig-lumin}d). A shock is launched 
when the inner core bounces upon reaching supra-nuclear density. 
As the shock passes through the neutrino trapping surface, i.e., the
``neutrinosphere'' at a density of $\rho\sim 10^{12}$~g/cm$^3$, 
protons liberated from nuclei by shock
heating rapidly capture electrons to produce a burst of $\nu_e$ on a 
timescale of $\sim10$~ms (see Fig.~\ref{Fig-lumin}a). The luminosity 
of this so-called neutronization $\nu_e$ burst is 
$\sim10^{53}$~erg/s and can provide a potential diagnostic of 
the neutrino mass hierarchy \cite{Kachelriess:2004ds} 
(see Sec.~\ref{sec-signal} for further discussion). The subsequent
neutrino emission has comparable luminosities for neutrinos and 
antineutrinos of all three flavors (see Fig.~\ref{Fig-lumin}). 
The timescale of $\sim10$~s for this emission is determined by 
neutrino diffusion out of the newly-formed proto-neutron star. 
 
The shock launched by core bounce is not energetic 
enough to break out of the outer core. It is stalled at $r_{\rm sh}\sim 100$~km 
and becomes an accretion shock through which matter can fall onto 
the proto-neutron star. During this accretion phase, the luminosities of 
$\nu_e$ and $\bar\nu_e$ are nearly twice as high as those of 
$\nu_{\mu(\tau)}$ and $\bar\nu_{\mu(\tau)}$ (see Fig.~\ref{Fig-lumin}b). 
This is because emission of $\nu_e$ and $\bar\nu_e$ is enhanced by
efficient charged-current reactions (dominantly $e^\pm$ capture on
free nucleons) in the extended region 
above the proto-neutron star while that of $\nu_{\mu(\tau)}$ and 
$\bar\nu_{\mu(\tau)}$ is dominated by diffusion out of the 
proto-neutron star. Absorption of some $\nu_e$ and $\bar\nu_e$ can
heat the material at $r<r_{\rm sh}$, thereby reviving the stalled shock
\cite{Bethe:1984ux}. However, recent studies suggest that this so-called
neutrino-driven explosion mechanism works robustly only for low-mass 
progenitors but must be combined with convection to deliver explosion
for higher-mass progenitors (see \cite{Janka:2012wk} for a review). In the 
latter case, multi-dimensional simulations are required. 

As an approximation to the effects of convection in multi-dimensional 
supernova models, neutrino heating in the region between the
neutrinosphere and the stalled shock is artificially enhanced in our
spherically symmetric model to trigger the explosion, thereby allowing 
us to study the long-term evolution of the proto-neutron star up to
a time post (core) bounce of $t_{\rm pb}\sim 10$~s. 
Once the shock is revived at $t_{\rm pb}\sim 350$~ms,
accretion of matter by the proto-neutron star quickly diminishes, 
resulting in a sharp drop of $\nu_e$ and $\bar\nu_e$ 
luminosities (see Fig.~\ref{Fig-lumin}b).
In the subsequent proto-neutron star cooling phase, 
the luminosities of $\nu_{\mu(\tau)}$ and $\bar\nu_{\mu(\tau)}$
become slightly higher than those of $\nu_e$ and $\bar\nu_e$,
because the former decouple from regions of higher temperature 
as reflected by their average energy (see Fig.~\ref{Fig-lumin}f). 
In general, the canonical average energy hierarchy of 
$\langle E_{\nu_e}\rangle <\langle E_{\bar\nu_e}\rangle <
\langle E_{\nu_{\mu(\tau)}}\rangle\approx 
\langle E_{\bar\nu_{\mu(\tau)}}\rangle$
holds throughout the accretion and cooling phases in our model
(see Figs.~\ref{Fig-lumin}e and f).

\begin{figure}
\includegraphics[width=0.9\columnwidth, angle=-90]{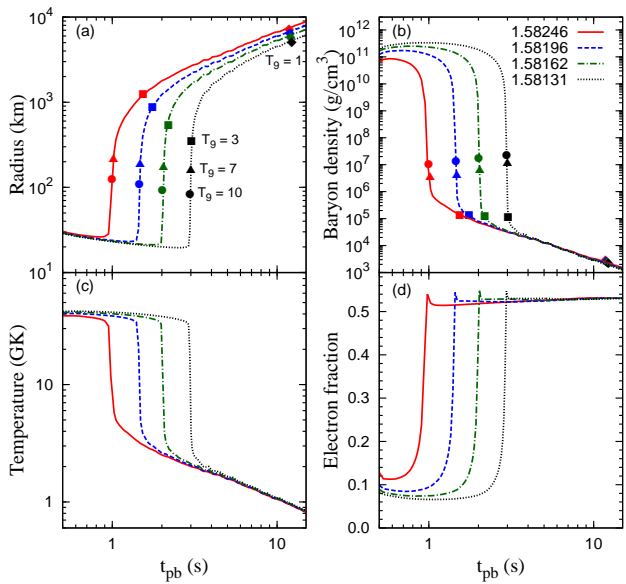}
\caption{Evolution of (a) radius, (b) density, (c) temperature, 
and (d) electron fraction for four mass elements in the neutrino-driven wind
of the $18\,M_\odot$ supernova model. The mass elements are
labeled by their enclosed baryonic masses in units of $M_\odot$.
The radius and density at which the temperature in units of GK
reaches $T_9=10$, 7, 3, and 1 are marked in panels (a) and (b).}
\label{Fig-tracer}
\end{figure}

During the cooling phase, $\nu_e$ and $\bar\nu_e$ continue to heat 
the material immediately outside the proto-neutron star, giving rise to
a matter outflow usually referred to as the neutrino-driven wind.
Specifically, free neutrons and protons in the wind material at high 
density and temperature can absorb $\nu_e$ and $\bar\nu_e$,
respectively, through
\begin{subequations}
\begin{align}
\nu_e+n &\to p+e^-,\label{eq-nuen}\\
\bar\nu_e+p &\to n+e^+.\label{eq-nuep}
\end{align}
\end{subequations}
This neutrino heating 
drives the wind to expand rapidly on timescales of $\sim 10$~ms
(see Fig.~\ref{Fig-tracer}). Under such conditions elements heavier than 
iron can form when the wind expands to low density and temperature.
A key parameter governing this nucleosynthesis is the electron fraction
$Y_e$, which is determined by the competition between 
reactions~(\ref{eq-nuen}) and (\ref{eq-nuep}). 
If $\nu_e$ and $\bar\nu_e$ had the same luminosities and energy
spectra, reaction~(\ref{eq-nuen}) would proceed faster than 
reaction~(\ref{eq-nuep}) because the former is favored by the 
neutron-proton mass difference $\Delta$ [see Eq.~(\ref{eq-crosection})].
In order to obtain a neutron-rich wind with $Y_e<0.5$ required for the 
rapid neutron-capture process (e.g., \cite{Arcones:2012wj}), the average  
energy of $\bar\nu_e$ must exceed that of $\nu_e$ by approximately 
$4\Delta$ with the same luminosity
for $\bar\nu_e$ and $\nu_e$ \cite{Qian:1996xt}.
While the luminosities are approximately the same, the average energy 
for $\bar\nu_e$ never exceeds that for $\nu_e$ by $4\Delta$ throughout
the cooling phase in our model. Consequently, the wind is proton rich 
as shown in Fig.~\ref{Fig-tracer}d for four selected mass elements.
In this case, a $\nu p$ process can occur \cite{Frohlich:2005ys,Pruet:2005qd,Huther:2013dza}.

The selected wind mass elements shown in Fig.~\ref{Fig-tracer}
enclose baryonic masses of 1.58246, 1.58196, 1.58162, and 1.58131 
in units of $M_\odot$, and are ejected from the proto-neutron star at
$t_{\rm pb}=0.840$, 1.253, 1.726, and 2.526~s, respectively.
The final abundances of nuclei produced by the $\nu p$ process
in these mass elements not only depend on their asymptotic $Y_e$ 
but also on the rate of reaction~(\ref{eq-nuep}) when their temperature 
evolves through the range of $1\lesssim T\lesssim 3$~GK 
\cite{Frohlich:2005ys,Pruet:2005qd}. As shown in Fig.~\ref{Fig-tracer}b, 
these mass elements stay in this temperature range for $\sim 10$~s. 
During this time, the neutrino luminosities change by an order of magnitude,
and so does the difference in average energy between $\bar\nu_e$ and 
$\bar\nu_{\mu(\tau)}$ (see Fig.~\ref{Fig-lumin}). We will show that
neutrino oscillations occur before the mass elements enter the above
temperature range and the results are extremely sensitive to the evolution 
of neutrino energy spectra. Therefore, we must conduct a comprehensive
study of neutrino oscillations for the entire cooling phase in order to 
examine their effects on nucleosynthesis in the neutrino-driven wind.

\section{Methodology of Calculating Neutrino Flavor Evolution}\label{sec-method}
We adopt the following neutrino mixing parameters:
$\Delta m_{21}^2=7.59\times 10^{-5}$~eV$^2$,
$|\Delta m_{31}^2|=2.43\times 10^{-3}$ eV$^2$, and
$\sin^2{2\theta_{12}}=0.87$ \cite{Nakamura:2010zzi}. 
Recent measurement of $\bar\nu_e$ disappearance at Daya Bay 
gave $\sin^2{2\theta_{13}}=0.092\pm 0.016\pm 0.005$ \cite{An:2012eh},
which corresponds to a central value of $\theta_{13}=0.15$.
This is somewhat larger than the value of $\theta_{13}=0.1$
assumed in \cite{Duan:2006an}, a major study on collective oscillations. 
We perform a full set of calculations using $\theta_{13}=0.1$ for 
comparison with this previous study, but also carry out additional
calculations using $\theta_{13}=0.15$. We find that the results
for these two values of $\theta_{13}$ agree within 5\%
(see Sec.~\ref{sec-tv13}). We consider both cases of
$\Delta m_{31}^2>0$ (normal mass hierarchy) and 
$\Delta m_{31}^2<0$ (inverted mass hierarchy). 
For easy separation of the normal and inverted mass hierarchies, we use 
a rotated flavor basis $(|\nu_e\rangle, |\nu_x\rangle, |\nu_y\rangle)^T=
R_{23}^{-1}(\theta_{23})(|\nu_e\rangle, |\nu_\mu\rangle, |\nu_\tau\rangle)^T$, 
where $R_{23}$ is the rotation matrix in the 2-3 subspace
\cite{Dasgupta:2007ws}. For all calculations we assume $\delta_{CP}=0$ 
(see \cite{Gava:2008rp} for discussion of generally small effects of 
$\delta_{CP}$ on supernova neutrino oscillations).

\begin{figure}
\includegraphics*[width=0.5\columnwidth, angle=-90]{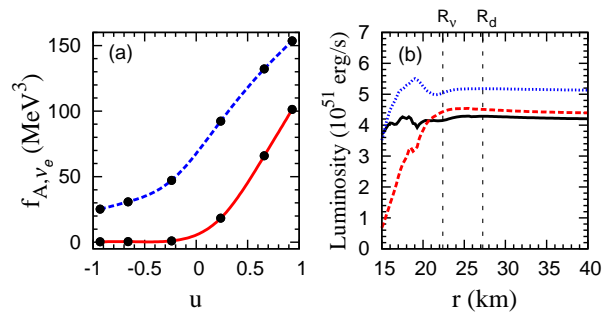} 
\caption{Neutrino decoupling in the $18\,M_\odot$ supernova model.
(a) Comparison of $\nu_e$ angular distributions at the conventional
neutrinosphere ($r=R_\nu$, blue dashed curve) and the decoupling
sphere ($r=R_d$, red solid curve) for $t_{\rm pb}=1.025$~s.
(b) Luminosities (corrected for
gravitational redshift) of $\nu_e$ (black solid curve), $\bar\nu_e$ 
(red dashed curve), and $\nu_{\mu(\tau)}$ (blue dotted curve)
as functions of radius for $t_{\rm pb}=1.025$~s, exhibiting constancy 
at $r>R_d$ to very good approximation. The luminosity of 
$\bar\nu_{\mu(\tau)}$ is essentially the same as that of $\nu_{\mu(\tau)}$.
\label{Fig-Rd313}}
\end{figure}

Studies of neutrino oscillations outside the proto-neutron star are usually 
carried out by adopting a neutrino emission model similar to the ``bulb model'' 
in \cite{Duan:2006an}, where all neutrinos are assumed to be free-streaming 
outward from a sharp neutrinosphere at $r=R_\nu$. The conventional 
neutrinosphere is defined as the surface outside which the neutrino optical depth
is 2/3. Consequently, a significant amount of scattering and emission still occurs
at $r=R_\nu$, giving rise to a significant neutrino flux that is propagating inward 
as shown in Fig.~\ref{Fig-Rd313}a for $\nu_e$ at $t_{\rm pb}=1.025$~s in our 
supernova model. In this figure and hereafter, we use $u\equiv\cos\theta$ to 
represent the angle of propagation $\theta$ with respect to the radial direction 
at radius $r$ ($u<0$ for inward-propagating neutrinos).
For our calculations of supernova neutrino flavor evolution, we start from
a decoupling sphere at $r=R_d$ where all inward-propagating neutrino fluxes are 
negligible ($\lesssim 2\%$ of the corresponding outward-propagating fluxes in 
general). As an example to justify our choice of $R_d$, we show the luminosities 
(corrected for gravitational redshift) for different neutrino flavors at 
$t_{\rm pb}=1.025$~s as functions of radius in Fig.~\ref{Fig-Rd313}b. 
It can be seen that all luminosities stay constant at $r>R_d$ to very good 
approximation.

We employ the neutrino energy and angular distributions 
$f_\nu(t_{\rm em},E,u_d,R_d)$ at $r=R_d$ (with $u_d\equiv\cos\theta_d$)
from our supernova model, which change significantly over $\sim 10$~s 
as illustrated by the evolution of average neutrino energies in 
Fig.~\ref{Fig-lumin}. We emphasize that it is important to use realistic 
neutrino energy and angular distributions at emission in calculating 
supernova neutrino flavor evolution, especially the collective oscillations.
In particular, we note that in contrast to the isotropic emission typically 
assumed in previous studies of collective oscillations, realistic neutrino 
angular distributions are forward peaked as shown for $\nu_e$ in 
Fig.~\ref{Fig-Rd313}a. The effects of $f_\nu(t_{\rm em},E,u_d,R_d)$ 
on collective neutrino oscillations will be discussed in Sec.~\ref{sec-multi}. 

In the absence of neutrino oscillations, the neutrino distributions at $r>R_d$
are given by
\begin{equation}\label{eq-fnu}
f_\nu(t_{\rm em},E,u,r)=f_\nu(t_{\rm em},E,u_d,R_d),
\end{equation} 
where $u$ and $u_d$ (see Fig.~\ref{Fig-sketch}) are related by
\begin{equation}\label{eq-uud}
u=\sqrt{1-(R_d/r)^2(1-u_d^2)}\,.
\end{equation}
The corresponding neutrino number density distributions per unit energy 
interval per unit solid angle at $r>R_d$ are given by
\begin{equation}\label{eq-nuden}
\frac{d^2n_\nu}{dEd\Omega}=\frac{E^2}{(2\pi)^3}f_\nu(t_{\rm em},E,u,r),
\end{equation}
where $d\Omega\equiv dud\phi$ with $\phi$ being the azimuthal angle.
In the above equation and elsewhere in the paper, natural units with 
$\hbar=c=1$ are used. 

The neutrino number density distributions in Eq.~(\ref{eq-nuden}) have 
azimuthal symmetry around the radial direction. We assume that this 
symmetry also applies to neutrino flavor evolution at $r>R_d$, 
where neutrinos experience forward scattering on other neutrinos and on 
electrons. The latter have a spherically symmetric number density profile 
$n_e(r)$ in our supernova model. 
Under the above assumption, the wave function 
$\psi_\nu(t_{\rm em},E,u,r)$ for a neutrino emitted with energy $E$ at time 
$t_{\rm em}$ and arriving at radius $r$ with a propagation angle specified
by $u$ satisfies a Schr{\"o}dinger-like equation,
\begin{equation}\label{eq-nuosct}
i\frac{d\psi_\nu}{dt}=
(H_{\rm v}+H_e+H_\nu)\psi_\nu(t_{\rm em},E,u,r),
\end{equation}
where $H_{\rm v}$, $H_e$, and $H_\nu$ are the effective Hamiltonians 
due to vacuum neutrino masses, neutrino-electron forward scattering 
\cite{Wolfenstein:1977ue}, and neutrino-neutrino forward scattering 
\cite{Fuller1987,Sigl:1992fn,Pantaleone:1992xh}, respectively. 
In our rotated flavor basis $(|\nu_e\rangle, |\nu_x\rangle, |\nu_y\rangle)^T$,
the wave function is $\psi_\nu(t_{\rm em},E,u,r)=(a_e, a_x, a_y)^T$, where
$a_e$, $a_x$, and $a_y$ are the amplitudes for being a $\nu_e$, $\nu_x$,
and $\nu_y$, respectively. In the same basis, the effective Hamiltonians are
\begin{subequations}\label{eq-Heff}
\begin{align}
H_{\rm v} & = U\frac{M^2}{2E}U^\dagger, \\
H_e & = \sqrt{2}G_F n_e(r){\rm diag}(1,0,0), \\
H_\nu & =\sqrt{2}G_F\sum_\alpha\int dE'd\Omega'(1-uu')
\left[\frac{d^2n_{\nu_\alpha}}{dE'd\Omega'}\right.\\
&\left.\times\rho_{\nu_\alpha}(t_{\rm em}',E',u',r)-
\frac{d^2n_{\bar \nu_\alpha}}{dE'd\Omega'}
\rho^*_{\bar \nu_\alpha}(t_{\rm em}',E',u',r)\right].\nonumber
\end{align}
\end{subequations}
In the above equations, $M={\rm diag}(m_1,m_2,m_3)$, 
$U=R_{13}(\theta_{13}) R_{12}(\theta_{12})$, 
$n_e=\rho Y_eN_A$ with $N_A$ being Avogadro's number,
$\rho_{\nu_\alpha}=\psi_{\nu_\alpha}\psi_{\nu_\alpha}^\dagger$,
and $\alpha=\{e,x,y\}$ denotes the initial neutrino flavor.

\begin{figure}
\includegraphics[width=0.6\columnwidth, angle=-90]{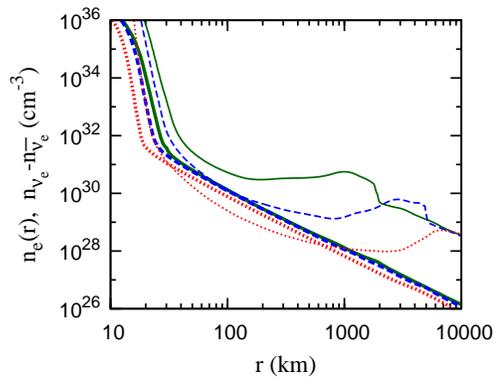}
\caption{Profiles of $n_e(r)$ (thin curves) and the net $\nu_e$ number density
$n_{\nu_e}(r)-n_{\bar\nu_e}(r)$ in the absence of neutrino oscillations
(thick curves) for $t_{\rm pb}\approx 0.6$ (green solid curve), 
1.0 (blue dashed curve), and 3.0~s (red dotted curve), respectively,
in the $18\,M_\odot$ supernova model.}
\label{Fig-density}
\end{figure}

The nonlinear coupling among all neutrinos introduced by $H_\nu$ 
can lead to collective oscillations (see \cite{Duan:2010bg} for a review and 
\cite{Raffelt:2011yb,Duan:2010bf,Chakraborty:2011nf,Chakraborty:2011gd,
Galais:2011gh,Wu:2011yi,Mirizzi:2011tu,Pehlivan:2011hp,Banerjee:2011fj,
Sarikas:2011am,Sarikas:2012vb,Saviano:2012yh,Cherry:2012zw,Volpe:2013uxl,
Raffelt:2013rqa,Raffelt:2013isa,Cherry:2013mv,Vlasenko:2013fja,Mirizzi:2013rla,
Duan:2013kba,Chakraborty:2014nma,Akhmedov:2014ssa,Vlasenko:2014bva} 
for more recent developments). 
To estimate the relative importance of $H_e$ and $H_\nu$ for our supernova
model, we show in Fig.~\ref{Fig-density} the profiles of $n_e(r)$ and the net 
$\nu_e$ number density $n_{\nu_e}(r)-n_{\bar\nu_e}(r)$ in the absence of 
neutrino oscillations for $t_{\rm pb}\approx 0.6$, 1.0, and 3.0~s, respectively.
It can be seen that $n_{\nu_e}-n_{\bar\nu_e}$ can exceed $n_e$ for
some radii only at $t_{\rm pb}\gtrsim 1.0$~s. Recent studies 
\cite{EstebanPretel:2008ni,Chakraborty:2011gd,Chakraborty:2014nma} 
suggest that for $n_{\nu_e}-n_{\bar\nu_e}\ll n_e$, collective oscillations are
suppressed due to large dispersion in $H_e$ for neutrinos with different 
propagation angles. Consequently, we expect that collective oscillations are 
suppressed for $t_{\rm pb}< 1.0$~s, i.e., during the accretion phase and 
the very early cooling phase in our supernova model. We focus our 
numerical calculations of neutrino flavor evolution on
the period of $0.6\lesssim t_{\rm pb}\lesssim 10.0$~s, 
during which collective oscillations might occur.
 
We note that although $f_\nu(t_{\rm em},E,u_d,R_d)$ and $n_e(r)$ change 
significantly over $0.6\lesssim t_{\rm pb}\lesssim 10.0$~s, they can be taken 
as fixed during the time between emission of a neutrino at $r=R_d$ and its 
arrival at $r\lesssim 500$~km, where $H_\nu$ might drive collective 
oscillations (see Fig.~\ref{Fig-density}). Therefore, in solving
Eq.~(\ref{eq-nuosct}), we first consider a snapshot of 
$f_\nu(t_{\rm em},E,u_d,R_d)$ and $n_e(r)$ for a specific $t_{\rm em}$,
and then use the corresponding $H_\nu$ and $H_e$ to evolve 
$\psi_\nu(t_{\rm em},E,u,r)$ at $r>R_d$. A total of $\approx 50$ snapshots 
are taken to cover $0.6\lesssim t_{\rm pb}\lesssim 10.0$~s. 
For each snapshot, we map $f_\nu(t_{\rm em},E,u_d,R_d)$ from our 
supernova model onto a grid of $E$ and $v\equiv u_d^2$.
As $u=\sqrt{1-(R_d/r)^2(1-v)}$ [see Eq.~(\ref{eq-uud})], the dependence on
$u$ is equivalent to that on $v$. Using $dt=dr/u$ (see Fig.~\ref{Fig-sketch}), 
we can rewrite Eq.~(\ref{eq-nuosct}) as
\begin{equation}\label{eq-nuoscr}
i\frac{d\psi_\nu}{dr}=\left(
\frac{H_{\rm v}+H_{\rm e}}{u}+H_\nu'\right)\psi_\nu(t_{\rm em},E,v,r),
\end{equation}
where
\begin{align}
&H_\nu' =\pi\sqrt{2}G_F\left(\frac{R_d}{r}\right)^2\sum_\alpha\int dE'dv'
\left(\frac{1}{uu'}-1\right)\times\\
&\left[\frac{d^2n_{\nu_\alpha}}{dE'd\Omega'}
\rho_{\nu_\alpha}(t_{\rm em},E',v',r)-
\frac{d^2n_{\bar \nu_\alpha}}{dE'd\Omega'}
\rho^*_{\bar \nu_\alpha}(t_{\rm em},E',v',r)\right].\nonumber
\end{align}
The evolution equation for $\psi_{\bar\nu}(t_{\rm em},E,v,r)$ can be 
obtained by the substitution $H_e\rightarrow -H_e$ and 
$H_\nu'\rightarrow -{(H_\nu')}^*$ in Eq.~(\ref{eq-nuoscr}). The results for
$\psi_\nu(t_{\rm em},E,v,r)$ and $\psi_{\bar\nu}(t_{\rm em},E,v,r)$ are
presented in Sec.~\ref{sec-nuosc}.

\section{Results on Collective Neutrino Oscillations}\label{sec-nuosc}
\begin{figure*}
\includegraphics*[width=0.7\columnwidth, angle=-90]{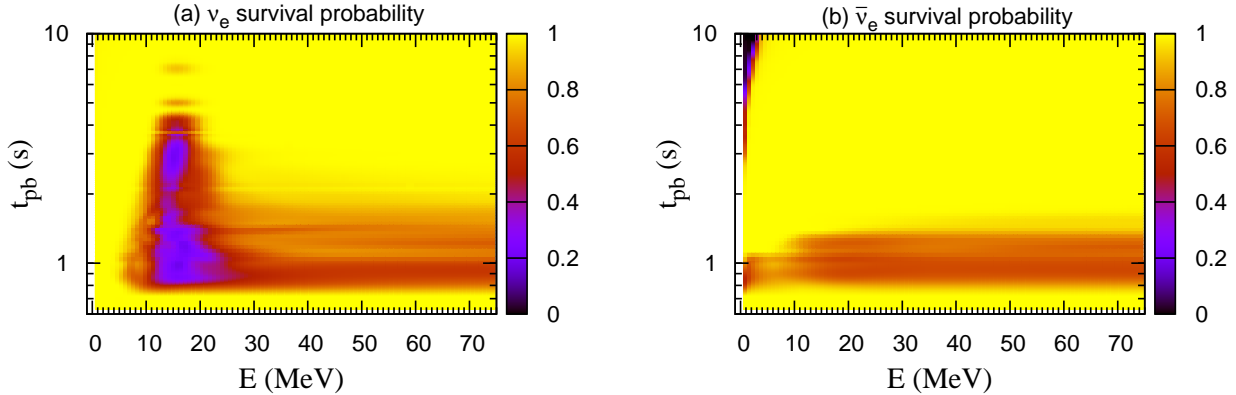}
\caption{The angle-averaged survival probabilities
(a) $\langle P_{\nu_e\nu_e}\rangle_v$ and 
(b) $\langle P_{\bar\nu_e\bar\nu_e}\rangle_v$ as functions of 
$E$ and emission time $t_{\rm em}$ (in terms of $t_{\rm pb}$)
at $r=500$ km.
\label{Fig-swap_r500}}
\end{figure*}

By solving Eq.~(\ref{eq-nuoscr}), we find that no significant neutrino 
flavor evolution occurs at $r\leq500$~km in our supernova model for 
the normal mass hierarchy, and that collective neutrino oscillations of 
particular interest to us have already ceased at $r=500$~km for the 
inverted mass hierarchy. We focus on the latter case and present the 
corresponding results at $r\leq 500$~km for 
$0.6\lesssim t_{\rm pb}\lesssim 10.0$~s in this section. 
We define the angle-averaged survival probability of $\nu_e$ as
\begin{subequations}
\begin{align}
\langle P_{\nu_e\nu_e}\rangle_v&\equiv
\frac{\int P_{\nu_e\nu_e}(t_{\rm em},E,v,r)
f_{\nu_e}(t_{\rm em},E,u,r)du}{\int f_{\nu_e}(t_{\rm em},E,u,r)du},\\
&=\frac{\int P_{\nu_e\nu_e}(t_{\rm em},E,v,r)
f_{\nu_e}(t_{\rm em},E,u_d,R_d)dv/u}{\int f_{\nu_e}(t_{\rm em},E,u_d,R_d)dv/u},
\end{align}
\end{subequations}
where we have used $f_{\nu_e}(t_{\rm em},E,u,r)=f_{\nu_e}(t_{\rm em},E,u_d,R_d)$
[Eq.~(\ref{eq-fnu})], $v=u_d^2$, and $u=\sqrt{1-(R_d/r)^2(1-v)}$ [Eq.~(\ref{eq-uud})]
to give the second expression. We show $\langle P_{\nu_e\nu_e}\rangle_v$ and 
$\langle P_{\bar\nu_e\bar\nu_e}\rangle_v$ as functions of $t_{\rm em}$ 
and $E$ for $r=500$~km in Fig.~\ref{Fig-swap_r500} and summarize
these results below:
\begin{enumerate}
\item For $t_{\rm pb}\lesssim 0.8$~s, collective oscillations are suppressed
by the large $H_e$ as expected;
\item For $0.8< t_{\rm pb}\lesssim 1.5$~s, 
significant flavor conversion occurs for neutrinos with $E\gtrsim 8$~MeV and 
for most antineutrinos;
\item For $1.5< t_{\rm pb}\lesssim 5.0$~s, only neutrinos with 
$10\lesssim E\lesssim 20$~MeV undergo collective oscillations;
\item For $t_{\rm pb}> 5.0$~s, collective oscillations are highly 
suppressed for all neutrinos and antineutrinos, and flavor conversion of 
low-energy $\bar\nu_e$ at these late times is driven by $H_e$ through
the adiabatic MSW effect.
\end{enumerate}

\begin{figure}
\includegraphics*[width=2.4\columnwidth, angle=-90]{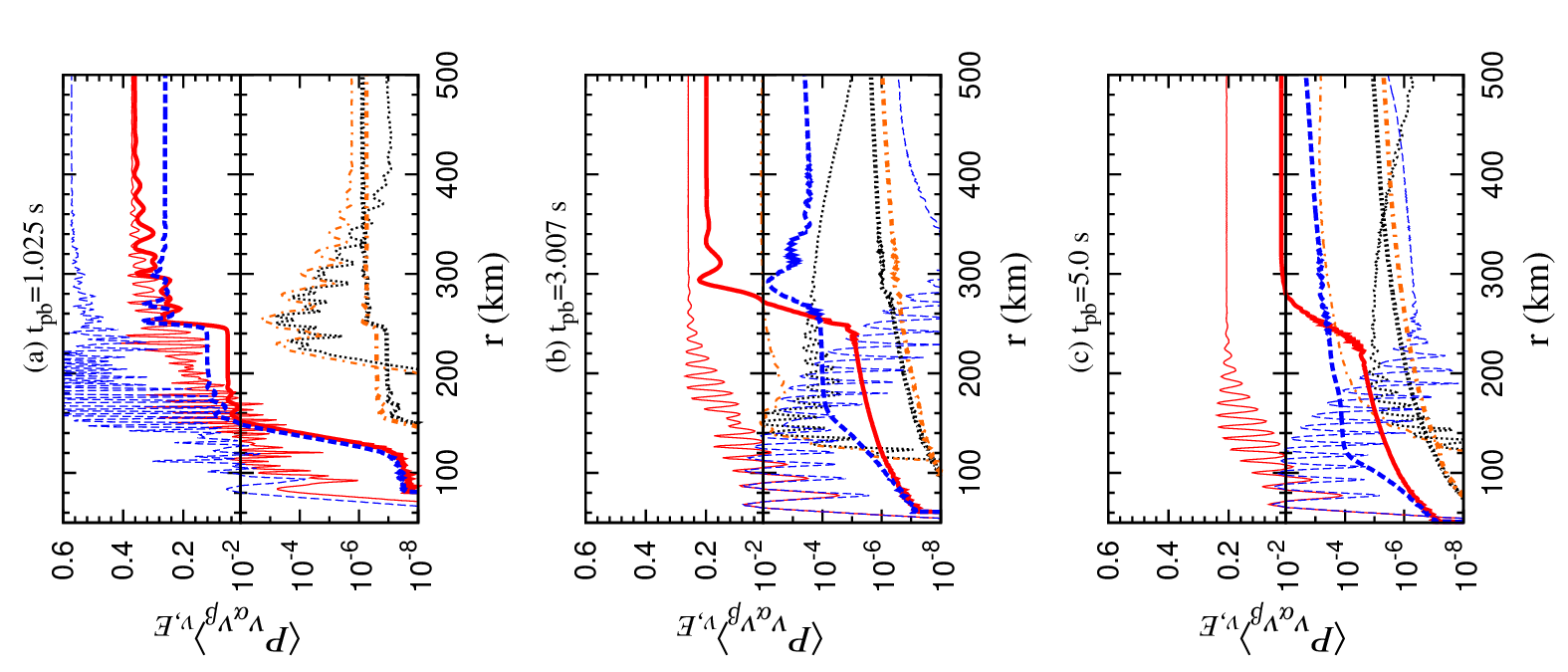}
\caption{Angle-and-energy-averaged conversion probabilities
$\left\langle P_{\nu_e\nu_x} \right\rangle_{v,E}$ (thick black dotted curve), 
$\left\langle P_{\nu_e\nu_y} \right\rangle_{v,E}$ (thick red solid curve),
$\left\langle P_{\bar\nu_e\bar\nu_x} \right\rangle_{v,E}$ 
(thick orange dash-dotted curve), and 
$\left\langle P_{\bar\nu_e\bar\nu_y} \right\rangle_{v,E}$ 
(thick blue dashed curve) as functions of radius calculated from
multi-angle simulations for $t_{\rm pb}=1.025$ (a), 3.007 (b), 
and 5.0~s (c). Similar quantities from single-angle calculations
are shown as the corresponding thin curves. \label{Fig-sumS}}
\end{figure}

The effective neutrino energy spectra at $r=500$~km can be obtained
from the survival probabilities shown in Fig.~\ref{Fig-swap_r500}.
In addition, it is important to examine the detailed neutrino flavor evolution 
at $r<500$~km so that its effects on physical processes at these radii 
can be assessed. We define the angle-and-energy-averaged probability 
for conversion of an initial $\nu_e$ into a $\nu_x$ as
\begin{subequations}
\begin{align}
&\langle P_{\nu_e\nu_x}\rangle_{v,E}\equiv
\frac{\int P_{\nu_e\nu_x}(t_{\rm em},E,v,r)
E^2f_{\nu_e}(t_{\rm em},E,u,r)dEdu}
{\int E^2f_{\nu_e}(t_{\rm em},E,u,r)dEdu},\\
&=\frac{\int P_{\nu_e\nu_x}(t_{\rm em},E,v,r)
E^2f_{\nu_e}(t_{\rm em},E,u_d,R_d)dEdv/u}
{\int E^2f_{\nu_e}(t_{\rm em},E,u_d,R_d)dEdv/u}.
\end{align}
\end{subequations}
The thick curves in Fig.~\ref{Fig-sumS} show 
$\langle P_{\nu_e\nu_x} \rangle_{v,E}$, 
$\langle P_{\nu_e\nu_y} \rangle_{v,E}$,
$\langle P_{\bar\nu_e\bar\nu_x} \rangle_{v,E}$, and 
$\langle P_{\bar\nu_e\bar\nu_y} \rangle_{v,E}$ 
as functions of radius for 
$t_{\rm pb}=1.025$, 3.007, and 5.0~s, respectively.

In general, the growth of $\langle P_{\nu_e\nu_y}\rangle_{v,E}$ 
($\langle P_{\bar\nu_e\bar\nu_y}\rangle_{v,E}$) corresponds to
flavor conversion between $\nu_e$ ($\bar\nu_e$) and 
$\nu_y$ ($\bar\nu_y$) in the 1-3 subspace associated with the larger 
mass squared-difference $|\Delta m_{31}^2|$, while that of
$\langle P_{\nu_e\nu_x}\rangle_{v,E}$
and $\langle P_{\bar\nu_e\bar\nu_x}\rangle_{v,E}$ corresponds to
flavor conversion associated with $\Delta m_{21}^2$. 
Because flavor instabilities, 
which mark the rapid growth of the flavor conversion probabilities,
are greatly suppressed by the multi-angle 
effects from both $H_e$ and $H_\nu$ 
(e.g., \cite{EstebanPretel:2008ni,Duan:2010bf,Banerjee:2011fj}),
large $e$-$y$ oscillations occur only at $r\sim 100$--300~km and 
$e$-$x$ oscillations are always negligible. Flavor evolution shown
in Fig.~\ref{Fig-sumS}a is representative of that for 
$0.8< t_{\rm pb}\lesssim 1.5$~s, when significant flavor conversion occurs
in both the neutrino and antineutrino sectors. 
In this case, there are two different flavor instabilities occurring at 
$r\approx 120$ and 240~km, respectively. The first instability induces 
more oscillations of antineutrinos, while the second affects neutrinos more.
In contrast, there is only one flavor instability affecting mostly neutrinos at 
$200\lesssim r\lesssim 250$~km for $1.5< t_{\rm pb}\lesssim 5$~s 
(see Figs.~\ref{Fig-sumS}b and \ref{Fig-sumS}c). For these later times,
the average conversion probability also grows more slowly to smaller 
values, and a smaller portion of the neutrino spectrum is affected 
as shown in Fig.~\ref{Fig-swap_r500}a.

To facilitate further discussion of the above results, we define the vacuum 
oscillation frequency $\omega\equiv|\Delta m_{31}^2|/(2E)$ for neutrinos 
and $\omega=-|\Delta m_{31}^2|/(2E)$ for antineutrinos. We also define
a normalized neutrino energy spectrum as a function of $\omega$,
\begin{equation}
g(\omega)=\frac{|\Delta m_{31}^2|}{2\omega^2}\times\begin{cases}
[g_{\nu_e}(E)-g_{\nu_y}(E)],&\mbox{for $\omega>0$},\\
[g_{\bar\nu_y}(E)-g_{\bar\nu_e}(E)],&\mbox{for $\omega<0$},
\end{cases}
\end{equation}
where
\begin{subequations}
\begin{align}
&g_{\nu_\alpha}(E)=\tilde{g}_{\nu_\alpha}(E)/G_\nu,\\
&\tilde{g}_{\nu_\alpha}(E)=\int E^2f_{\nu_\alpha}(t_{\rm em},E,u_d,R_d)u_ddu_d,\\
&G_\nu=\int[\tilde{g}_{\nu_e}(E)-\tilde{g}_{\bar\nu_e}(E)-\tilde{g}_{\nu_y}(E)
+\tilde{g}_{\bar\nu_y}(E)]dE.
\end{align}
\end{subequations}
In the above equations, $\tilde{g}_{\nu_\alpha}(E)$, and hence, 
$G_\nu$, $g_{\nu_\alpha}(E)$, and $g(\omega)$, depend on $t_{\rm em}$. 
This dependence is suppressed for simplicity. Note that 
$f_{\nu_y}(t_{\rm em},E,u_d,R_d)=f_{\nu_{\mu(\tau)}}(t_{\rm em},E,u_d,R_d)$
and $f_{\bar\nu_y}(t_{\rm em},E,u_d,R_d)=f_{\bar\nu_{\mu(\tau)}}(t_{\rm em},E,u_d,R_d)$
in our supernova model. It is also useful to introduce two effective potentials
\begin{subequations}
\begin{align}
\lambda(r)&=\sqrt{2}G_F n_e(r)\frac{R_d^2}{2r^2},\\
\mu(r)& =\sqrt{2}G_F\left(\frac{G_{\nu}}{2\pi^2}\right)\frac{R_d^4}{4r^4},
\end{align}
\end{subequations}
which approximately represent the differences in $H_e$ and $H_\nu$ 
among neutrinos with different propagation angles, and can be used to measure 
the so-called ``multi-angle'' effects on collective oscillations \cite{Banerjee:2011fj}. 
As the net neutrino number density at radius $r$ is
$\sim n_{\nu_e}-n_{\bar\nu_e}\sim [G_\nu/(2\pi^2)]R_d^2/(2r^2)$,
comparing $\lambda(r)$ and $\mu(r)$ is roughly equivalent to comparing 
$n_e$ and $n_{\nu_e}-n_{\bar\nu_e}$ (see Fig.~\ref{Fig-density}).

\subsection{Multi-Angle Effects on Flavor Evolution}\label{sec-multi}
The above results on collective neutrino oscillations are obtained 
from the so-called ``multi-angle'' simulations in contrast to 
the ``single-angle'' approximation, which assumes that neutrino flavor 
evolution is independent of the propagation angle (e.g., \cite{Duan:2006an}). 
The single-angle approximation was widely used in the literature to facilitate
analytical understanding of collective oscillations. 
In some cases, the results from this 
approximation qualitatively agree with those from multi-angle simulations.
Examples include cases where fluxes of  $\nu_{\mu(\tau)}$ and 
$\bar\nu_{\mu(\tau)}$ are significantly smaller than those of $\nu_e$ and 
$\bar\nu_e$ \cite{Duan:2006an,Fogli:2007bk,Wu:2011yi} and the case of 
the neutronization burst of an O-Ne-Mg core-collapse supernova 
\cite{Cherry:2010yc,Cherry:2011fm,Cherry:2011fn}. However, for
neutrino energy spectra representative of the cooling phase,
the results from multi-angle simulations are typically very different from 
those obtained with the single-angle approximation
\cite{Dasgupta:2009mg,Duan:2010bf,Mirizzi:2010uz}.
Below we compare the multi-angle and single-angle results for our 
supernova model.

We perform single-angle calculations assuming that
neutrinos are emitted uniformly within the forward $2\pi$ solid angle
at $r=R_d$ with the same total fluxes and energy spectra as given 
by our supernova model and that neutrinos with the same energy 
undergo the same flavor evolution at $r>R_d$ as those propagating 
radially. The resulting energy-averaged conversion probabilities 
$\langle P_{\nu_e\nu_x} \rangle_E$, 
$\langle P_{\nu_e\nu_y} \rangle_E$,
$\langle P_{\bar\nu_e\bar\nu_x} \rangle_E$, and 
$\langle P_{\bar\nu_e\bar\nu_y} \rangle_E$ as functions of radius 
are shown as thin curves in Fig.~\ref{Fig-sumS} for 
$t_{\rm pb}=1.025$, 3.007, and 5.0~s, respectively.
It can be seen that there is rapid growth of 
$\langle P_{\nu_e\nu_y} \rangle_E$ and
$\langle P_{\bar\nu_e\bar\nu_y} \rangle_E$ at $r<100$~km in all cases. 
This early onset of flavor oscillations is due to a flavor instability that 
occurs in an isotropic environment when there is multiple spectral crossings 
corresponding to $g(\omega)=0$ even for arbitrarily large electron
and/or neutrino density \cite{Dasgupta:2009mg}.
This instability also triggers the onset of $e$-$x$ conversion in the 
region of collective oscillations \cite{Friedland:2010sc,Dasgupta:2010cd}.
In contrast, neutrinos with different propagation angles experience different 
histories of $H_e$ and $H_\nu$ in multi-angle calculations. This greatly 
suppresses flavor instabilities, with large $e$-$y$ oscillations occurring only 
at $r\sim 100$--300~km and $e$-$x$ oscillations being always negligible
(see Fig.~\ref{Fig-sumS}).

\begin{figure}
\includegraphics[width=2.4\columnwidth, angle=-90]{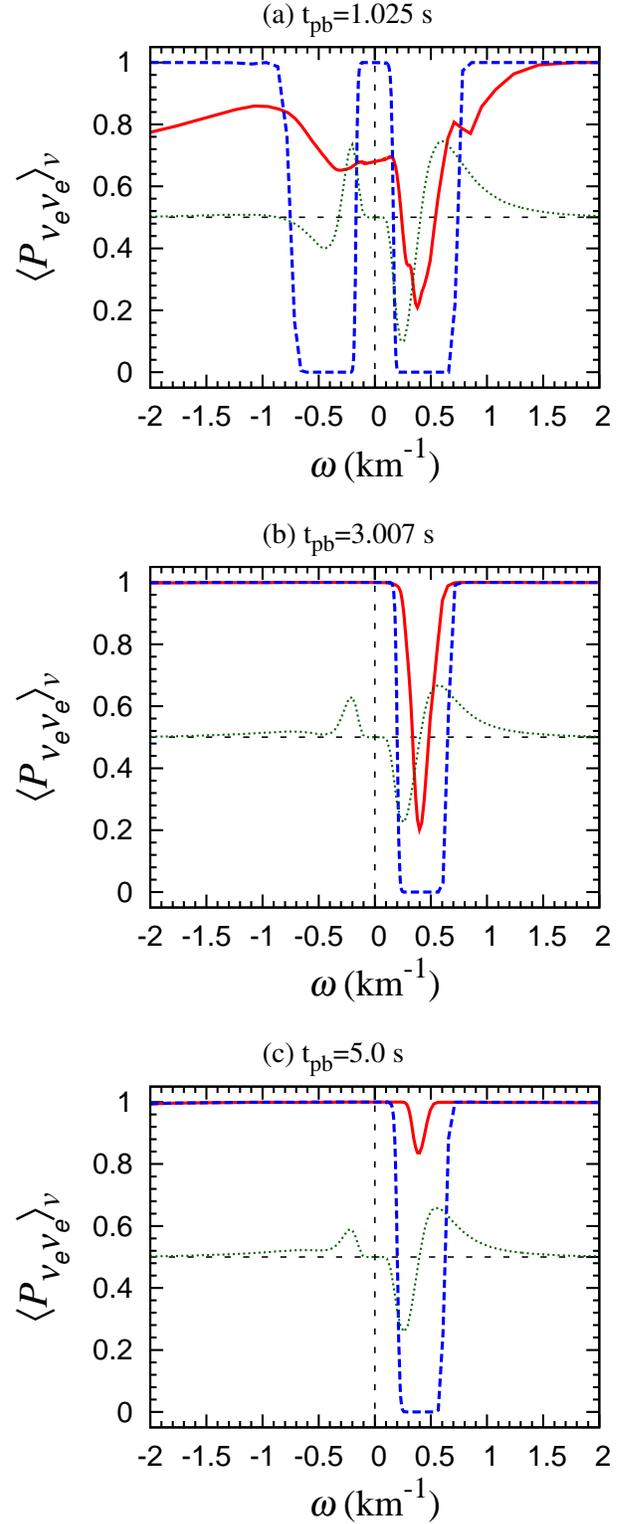}
\caption{Angle-averaged survival probabilities
$\langle P_{\nu_e\nu_e}\rangle_v$ ($\omega>0$) and 
$\langle P_{\bar\nu_e\bar\nu_e}\rangle_v$ ($\omega<0$) at $r=500$~km 
as functions of $\omega$ calculated from multi-angle simulations 
(red solid curve) for $t_{\rm pb}=1.025$ (a), 3.007 (b), and 5.0~s (c).
The corresponding single-angle results for $1- P_{\nu_e\nu_y}$ and
$1- P_{\bar\nu_e\bar\nu_y}$ are shown as blue dashed curves. 
The green dotted curves give the function $g(\omega)/15+0.5$.\label{Fig-swap}}
\end{figure}

The most distinct feature of collective oscillations is that $\nu_e$ and $\bar\nu_e$ 
can swap part of their spectra with $\nu_{\mu,\tau}$ and $\bar\nu_{\mu,\tau}$
\cite{Duan:2006an,Fogli:2007bk,Dasgupta:2009mg,Friedland:2010sc,Mirizzi:2010uz}.
Such spectral splits or swaps are best illustrated by the probabilities
$1-P_{\nu_e\nu_y}$ and $1- P_{\bar\nu_e\bar\nu_y}$ at $r=500$~km as functions 
of $\omega$ obtained from the single-angle calculations, which are shown as 
the blue dashed curve in Fig.~\ref{Fig-swap} for 
$t_{\rm pb}=1.025$, 3.007, and 5.0~s, respectively. The function
$g(\omega)/15+0.5$ is shown as the green dotted curve in the same figure and
indicates that spectral splits or swaps could form around the ``positive'' spectral 
crossings corresponding to $g(\omega)=0$ and $dg/d\omega>0$ under the 
single-angle approximation \cite{Dasgupta:2009mg}.
However, because neutrinos with different propagation angles experience
different $H_\nu$ in multi-angle calculations, different parts of their energy 
spectra are in resonance when flavor instabilities or large-scale flavor 
oscillations occur (e.g., \cite{Wu:2011yi}).
Consequently, the splits in their energy spectra are generally smoothed out 
when averaged over the propagation angle. This can be seen from
the angle-averaged survival probabilities $\langle P_{\nu_e\nu_e}\rangle_v$ 
and $\langle P_{\bar\nu_e\bar\nu_e}\rangle_v$ at $r=500$~km as functions of 
$\omega$, which are shown as the red solid curve in Fig.~\ref{Fig-swap}.
Note that $\langle P_{\nu_e\nu_e}\rangle_v\approx 1-
\langle P_{\nu_e\nu_y}\rangle_v$ and 
$\langle P_{\bar\nu_e\bar\nu_e}\rangle_v\approx 1-
\langle P_{\bar\nu_e\bar\nu_y}\rangle_v $ because 
$\langle P_{\nu_e\nu_x}\rangle_v$ and $\langle P_{\bar\nu_e\bar\nu_x}\rangle_v$
are negligible in multi-angle calculations (see Fig.~\ref{Fig-sumS}).

We also note that flavor conversion of antineutrinos occurs only when 
there is an excess of $\bar\nu_e$ over $\bar\nu_y$, i.e., $g(\omega)<0$,
for some range of $\omega<0$. This is demonstrated in Fig.~\ref{Fig-swap}a
for $t_{\rm pb}=1.025$~s with $g(\omega)<0$ for 
$-0.9\lesssim\omega\lesssim-0.3$~km$^{-1}$
($7\lesssim E\lesssim 19$~MeV),
which is representative of the early deleptonization phase during protoneutron
star evolution. At later times, the energy spectra of $\bar\nu_e$ and 
$\bar\nu_y$ become similar but the luminosity of $\bar\nu_y$ remains 
higher than that of $\bar\nu_e$. This results in $g(\omega)>0$ for all 
$\omega<0$ as shown for $t_{\rm pb}=3.007$ and 5.0~s in 
Figs.~\ref{Fig-swap}b and \ref{Fig-swap}c, respectively. 
There is only one positive spectral crossing at $\omega>0$, 
i.e., in the neutrino sector, at these later times. Consequently, 
only the spectra of neutrinos are affected by collective oscillations 
(see Figs.~\ref{Fig-swap}b and \ref{Fig-swap}c).

\begin{figure}
\begin{tabular}{c}
\includegraphics*[width=60mm, angle=-90]{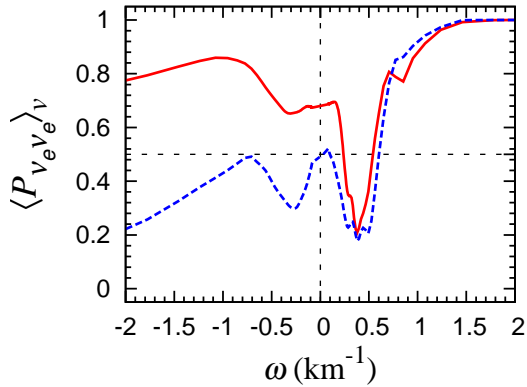}
\end{tabular}
\caption{Angle-averaged survival probabilities 
$\langle P_{\nu_e\nu_e}\rangle_v$ ($\omega>0$) and 
$\langle P_{\bar\nu_e\bar\nu_e}\rangle_v$ ($\omega<0$)
as functions of $\omega$
at $r=500$~km for $t_{\rm pb}=1.025$~s obtained from multi-angle 
simulations assuming isotropic neutrino emission 
(blue dashed curve).
The red solid curve (same as in Fig.~\ref{Fig-swap}a) is 
calculated with the forward-peaked neutrino angular distributions 
in our supernova model and is shown for comparison.
\label{Fig-swap_iso}}
\end{figure}

To conclude this subsection,
we examine the effects of neutrino angular distributions
on flavor evolution in supernovae using multi-angle simulations.
As shown in Fig.~\ref{Fig-Rd313}, the neutrino distributions 
$f(t_{\rm em},E,u_d,R_d)$ at the decoupling sphere are strongly 
forward-peaked instead of being isotropic as often assumed in earlier 
studies of collective oscillations. Compared to a physical
forward-peaked neutrino angular distribution, the crude assumption 
of isotropic neutrino emission leads to artificially larger $H_\nu$ for 
any specific neutrino trajectory because of the larger contributions 
from the more tangentially-emitted neutrinos.
In Fig.~\ref{Fig-swap_iso}, we show the angle-averaged
survival probability $\langle P_{\nu_e\nu_e}\rangle_v$ as a function
of $\omega$ at $t_{\rm pb}=1.025$~s (blue dashed curve) obtained 
from multi-angle calculations with the same total neutrino fluxes and 
energy spectra as given by our supernova model but
assuming isotropic neutrino emission. 
Compared with the result based on the neutrino angular distributions 
in our supernova model (red solid curve), the unphysical isotropic 
angular distribution gives much smaller survival probabilities for 
antineutrinos ($\omega<0$). 
It also causes the onset of flavor oscillations to occur 
at smaller radii. We emphasize that it requires not only multi-angle 
simulations, but also the use of proper neutrino angular distributions
to accurately treat collective neutrino oscillations in supernovae.

\subsection{Flavor Instabilities and Effects of $n_e$}\label{sec-ne}
\begin{figure}
\includegraphics*[width=0.9\columnwidth, angle=-90]{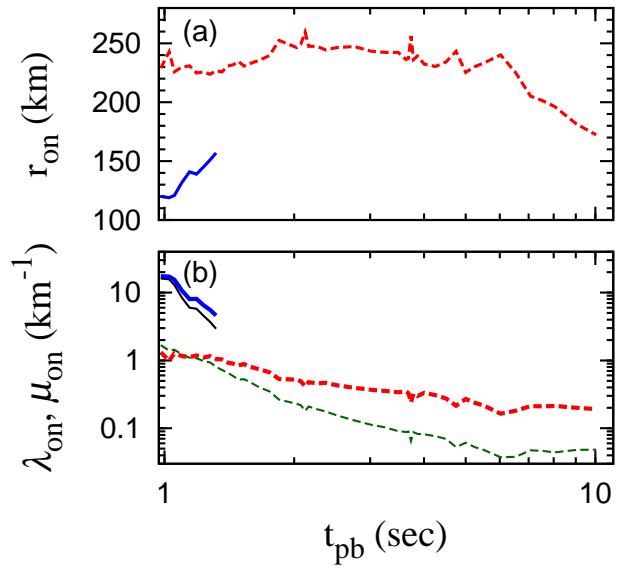}
\caption{(a) Onset radii $r_{\rm on}$ for two flavor instabilities as functions of 
time. (b) Corresponding $\lambda_{\rm on}=\lambda(r_{\rm on})$ (thin curves) 
and $\mu_{\rm on}=\mu(r_{\rm on})$ (thick curves) as functions of time.
The instability occurring at smaller radii (solid curves) only exists for
$t_{\rm pb}\sim 1$~s while that occurring at larger radii (dashed curves) exists
for $t_{\rm pb}\gtrsim 1$~s.
\label{Fig-onset}}
\end{figure}

As mentioned above, there are two different flavor instabilities occurring at 
$r\approx 120$ and 240~km, respectively, for collective neutrino oscillations
shown in Fig.~\ref{Fig-sumS}a, which are representative of the epoch of
$0.8< t_{\rm pb}\lesssim 1.5$~s. In contrast, there is only one flavor 
instability occurring at $200\lesssim r\lesssim 250$~km for 
$1.5< t_{\rm pb}\lesssim 5$~s (see Figs.~\ref{Fig-sumS}b and \ref{Fig-sumS}c). 
Using the detailed results from our multi-angle simulations, we show the radius
$r_{\rm on}$ for the onset of either instability as a function of time in
Fig.~\ref{Fig-onset}a. The corresponding 
$\lambda_{\rm on}=\lambda(r_{\rm on})$ and $\mu_{\rm on}=\mu(r_{\rm on})$ 
are shown in Fig. \ref{Fig-onset}b. The instability occurring at smaller radii is
tied to the substantial excess of $\bar\nu_e$ over $\bar\nu_y$ for some energy
range [e.g., $g(\omega)<0$ for $-0.9\lesssim\omega\lesssim-0.3$~km$^{-1}$ in 
Fig. \ref{Fig-swap}a] characteristic of the early deleptonization epoch at 
$t_{\rm pb}\sim1.0$~s. The corresponding values of
$\lambda_{\rm on}\sim\mu_{\rm on}\sim10$ km$^{-1}$ are much larger than 
the typical spread in $\omega$ of $\Delta\omega\sim 0.6$~km$^{-1}$. 
In contrast, the instability occurring at larger radii exists for 
$t_{\rm pb}\gtrsim1.0$~s because there is always an excess of $\nu_e$ over
$\nu_y$ for some energy range (see Figs.~\ref{Fig-swap}a and \ref{Fig-swap}b).
This instability generally corresponds to $\mu_{\rm on}\sim\Delta\omega$.

\begin{figure}
\includegraphics*[width=0.55\columnwidth, angle=-90]{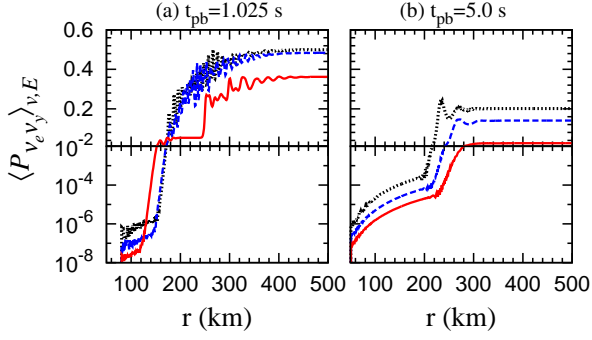}
\caption{Comparison of $\langle P_{\nu_e\nu_y}\rangle_{v,E}$ calculated
for $n_e(r)$ in our supernova model (red solid curves), $n_e'(r)=0.5n_e(r)$
(blue dashed curves), and $n_e''(r)=0.2n_e(r)$ (black dotted curves) for 
$t_{\rm pb}=1.025$ (a) and 5.0~s (b), respectively.
\label{Fig-sumS_ne}}
\end{figure}

As mentioned at the beginning of Sec.~\ref{sec-nuosc}, there are no significant
collective oscillations at $t_{\rm pb}>5$~s. This is puzzling because there is 
still a flavor instability occurring at $r\sim 200$~km for such times (see 
Fig.~\ref{Fig-onset}). We show below that this instability is suppressed by the 
effects of $n_e$, and therefore, fails to cause significant collective oscillations
at $t_{\rm pb}>5$~s. We perform multi-angle simulations with modified profiles 
of electron number density $n_e'(r)=0.5n_e(r)$ and $n_e''(r)=0.2n_e(r)$ at 
$r>R_d$, respectively, for $t_{\rm pb}=5$~s. We compare the corresponding
$\langle P_{\nu_e\nu_y}\rangle_{v,E}$ as functions of radius with the results 
calculated for the unmodified $n_e(r)$ in Fig.~\ref{Fig-sumS_ne}b. 
It can be seen that as $n_e(r)$ is reduced to $n_e'(r)$ and then to $n_e''(r)$,
the onset of flavor instability is pushed to smaller and smaller radii and its 
growth causes more and more flavor conversion. Therefore, collective
oscillations are suppressed by larger $n_e$. This generally holds for 
most of the cooling phase when there is only one flavor instability at 
$r\sim 200$~km. We also note that larger $n_e$ decreases the local 
effective mixing angle, which tends to reduce flavor conversion similar to 
the case of a small vacuum mixing angle.

\begin{figure}
\includegraphics*[width=0.53\columnwidth, angle=-90]{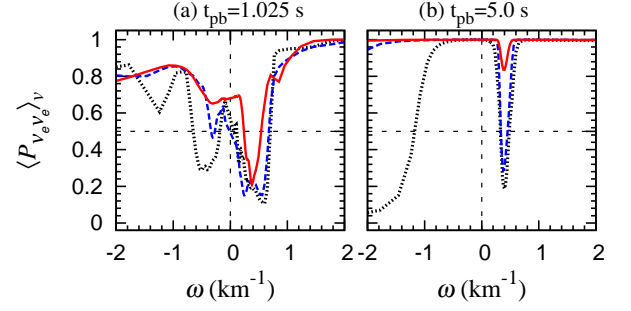}
\caption{Comparison of $\langle P_{\nu_e\nu_e}\rangle_v$ ($\omega>0$) 
and $\langle P_{\bar\nu_e\bar\nu_e}\rangle_v$ ($\omega<0$) at 
$r=500$~km calculated for $n_e(r)$ in our supernova model 
(red solid curves), $n_e'(r)=0.5n_e(r)$ (blue dashed curves), and 
$n_e''(r)=0.2n_e(r)$ (black dotted curves) for 
$t_{\rm pb}=1.025$ (a) and 5.0~s (b), respectively.
Note that in (b), the $\bar\nu_e$ flavor conversion
for $\omega\lesssim -1$~km$^{-1}$ is due to the 
MSW effect for the reduced $n_e''(r)$.
\label{Fig-swap_ne}}
\end{figure}

For completeness, we compare $\langle P_{\nu_e\nu_y}\rangle_{v,E}$ 
calculated for $n_e(r)$, $n_e'(r)$, and $n_e''(r)$
as functions of radius for $t_{\rm pb}=1.025$~s in Fig.~\ref{Fig-sumS_ne}a.
It can be seen that in the case of $n_e(r)$, the flavor instability at smaller 
radii stops growing when flavor conversion is still small and a second 
instability clearly occurs at larger radii. As $n_e(r)$ is reduced to $n_e'(r)$ 
and $n_e''(r)$, the ``first'' instability at $r\approx 150$~km grows to cause 
large flavor conversion and a ``second'' instability can no longer be 
identified clearly. We also compare $\langle P_{\nu_e\nu_e}\rangle_v$
and $\langle P_{\bar\nu_e\bar\nu_e}\rangle_v$ at $r=500$~km
calculated for $n_e(r)$, $n_e'(r)$, and $n_e''(r)$ as functions of $\omega$
for $t_{\rm pb}=1.025$ and 5.0~s in Figs.~\ref{Fig-swap_ne}a and 
\ref{Fig-swap_ne}b, respectively. It can be seen that as $n_e(r)$ is 
reduced to $n_e'(r)$ and then to $n_e''(r)$, features of spectral
swaps are increasingly sharpened and approach closer and closer to
the corresponding results for the single-angle approximation shown in
Figs.~\ref{Fig-swap}a and \ref{Fig-swap}c.

\begin{figure}
\includegraphics*[width=0.5\columnwidth, angle=-90]{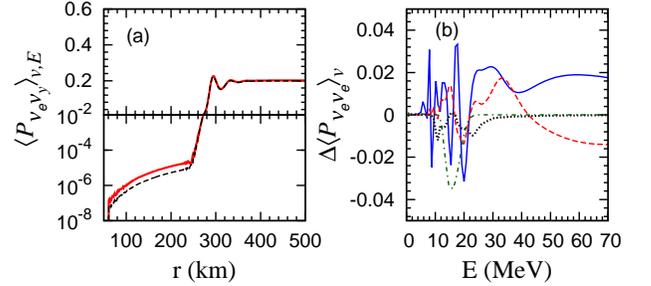}
\caption{(a) Comparison of $\langle P_{\nu_e\nu_y}\rangle_{v,E}$ 
for $\theta_{13}=0.15$ (red solid curve) and $0.1$ (black dashed curve)
as functions of radius for $t_{\rm pb}=3.007$~s.
(b) Difference in survival probability $\Delta\langle P_{\nu_e\nu_e}\rangle_v
\equiv\langle P_{\nu_e\nu_e}(\theta_{13}=0.15)\rangle_v
-\langle P_{\nu_e\nu_e}(\theta_{13}=0.1)\rangle_v$ at $r=500$~km
as functions of neutrino energy for $t_{\rm pb}=1.025$ (blue solid curve), 
1.401 (red dashed curve), 3.007 (black dotted curve), and 5.0~s 
(green dot-dashed curve), respectively.\label{Fig-tv}} 
\end{figure} 
\subsection{Results for Measured $\theta_{13}$\label{sec-tv13}}
Recent measurement of $\bar\nu_e$ disappearance by the Daya Bay 
experiment gave $\sin^2{2\theta_{13}}=0.092\pm 0.016\pm 0.005$ 
\cite{An:2012eh}, which corresponds to a central value of
$\theta_{13}=0.15$ that is somewhat larger than the value of 
$\theta_{13}=0.1$ adopted in the above calculations of neutrino 
oscillations. We have performed additional simulations for 
$\theta_{13}=0.15$ and confirm that our results on neutrino 
oscillations do not change significantly.
As an example, we compare $\langle P_{\nu_e\nu_y}\rangle_{v,E}$
for $\theta_{13}=0.15$ (red solid curve) and 0.1 (black dashed curve)
as functions of radius for $t_{\rm pb}=3.007$~s in Fig.~\ref{Fig-tv}a.
It can be seen that the flavor instability occurs at 
the same place and subsequent flavor evolution is identical
for both cases. The only change occurs before the onset of
the flavor instability, during which time
$\langle P_{\nu_e\nu_y}\rangle_{v,E}$ is slightly larger for
$\theta_{13}=0.15$ but is very small anyway.
We also show
$\Delta\langle P_{\nu_e\nu_e}\rangle_v\equiv
\langle P_{\nu_e\nu_e}(\theta_{13}=0.15)\rangle_v-
\langle P_{\nu_e\nu_e}(\theta_{13}=0.1)\rangle_v$ at $r=500$~km
as functions of neutrino energy for $t_{\rm pb}=1.025$, 1.401, 3.007,
and 5.0~s, respectively, in Fig.~\ref{Fig-tv}b.
It can be seen that $|\Delta\langle P_{\nu_e\nu_e}\rangle_v|$ 
is $\lesssim 4\%$ over the entire energy range in all cases.
We conclude that our results on neutrino oscillations are insensitive
to the exact value of $\theta_{13}$ and will use those calculated for
$\theta_{13}=0.1$ to examine the effects on supernova 
nucleosynthesis and neutrino signals.

\section{Effects of Neutrino Oscillations on Nucleosynthesis}

\label{sec-nucleo}

Neutrinos play important roles in supernova nucleosynthesis in
several major ways and in different locations.  For the
neutrino-driven wind (see Sec.~\ref{sec-model})
where initially matter is dominantly composed of free nucleons,
neutrino reactions~(\ref{eq-nuen}) and (\ref{eq-nuep}) set the
electron fraction $Y_e$, which is a crucial parameter governing the
nucleosynthesis in the ejecta.  For ejecta with $Y_e>0.5$ such as in
our supernova model (see Fig.~\ref{Fig-tracer}), a $\nu p$ process
occurs to produce heavy nuclei during expansion of the mass
elements~\cite{Frohlich:2005ys,Huther:2013dza}. This process
requires significant $\bar\nu_e$ absorption on protons when matter
evolves through the temperature range of $T=1$--3~GK.
In the outer envelope of the star, the interaction of
neutrinos with pre-existing nuclei can drive several nucleosynthesis
processes including the $\nu$ process
\cite{1990ApJ...356..272W,Heger:2003mm} 
and other neutrino-induced nucleosynthesis
\cite{1988PhRvL..61.2038E,1990ApJ...356..272W,Banerjee:2011zm,2013PhRvL.110n1101B}.
In the $\nu$ process, neutrinos can directly transform by
charged-current reactions abundant nuclear species into less
abundant neighboring nuclei. For example, 
$^{138}$La and $^{180}$Ta are known to
be produced by this mechanism from the abundant $^{138}$Ba and
$^{180}$Hf~\cite{Heger:2003mm,Byelikov.Adachi.ea:2007}. In
addition, neutral-current processes excite abundant nuclei to states
above particle emission. The decay of these nuclei and subsequent capture of the
produced protons, neutrons and/or $\alpha$ particles contribute to the
production of several nuclei including $^{7}$Li, $^{11}$B, and
$^{19}$F. The production of $^{19}$F is mainly due to neutral-current processes
\cite{Heger:2003mm} that are not affected by
neutrino oscillations. The situation is different for the other
species as discussed below.

In the helium shell, neutrinos interact with $^4$He through
the charged-current reactions
\begin{subequations}
\begin{align}
\nu_e+{^4{\rm He}}&\to{^3{\rm He}}+p+e^-,\label{eq-nuap}\\
\bar\nu_e+{^4{\rm He}}&\to{^3{\rm H}}+n+e^+.\label{eq-nuan}
\end{align}
\end{subequations}
The $^3$He and $^3$H produced by the above reactions
and by neutral-current spallation reactions on $^{4}$He are important to
the production of light nuclei such as $^7$Li and $^{11}$B in 
the $\nu$ process through the subsequent reactions
$^3$He$(\alpha,\gamma)^7$Be$(e^+\nu_e)^7$Li and
$^3$H$(\alpha,\gamma)^7$Li$(\alpha,\gamma)^{11}$B.
In addition, the neutrons produced by reaction~(\ref{eq-nuan}) may lead to 
a possible neutrino-induced $r$ process~\cite{1988PhRvL..61.2038E,Banerjee:2011zm}.
In the O/Ne shell, the production of $^{138}$La and $^{180}$Ta is predominantly
determined by the charged-current reactions:
\begin{subequations}
\begin{align}
\nu_e+{^{138}{\rm Ba}}&\to{^{138}{\rm La}}+e^-,\label{eq-nuBa}\\
\nu_e+{^{180}{\rm Hf}}&\to{^{180}{\rm Ta}}+e^-.\label{eq-nuHf}
\end{align}
\end{subequations} 
For the above cases, neutrino oscillations can affect
nucleosynthesis by changing the energy spectra of $\nu_e$ and
$\bar\nu_e$, and hence, the rates of charged-current $\nu_e$ and
$\bar\nu_e$ reactions. Many previous studies have discussed
the effects of neutrino oscillations on nucleosynthesis in the neutrino-driven wind
(e.g., \cite{MartinezPinedo:2011br,Duan:2010af}) and on the production ratio 
of $^7$Li to $^{11}$B \cite{Yoshida:2006sk}, using time-independent
parametrized neutrino spectra and/or ejecta trajectories.  Below we
apply our results of collective neutrino oscillations, consistently
calculated with realistic neutrino emission spectra in a dynamically
changing supernova environment as discussed in Sec.~\ref{sec-nuosc}, to
examine the effects of oscillations on the $\nu p$ process in the
neutrino-driven wind, on neutrino-induced nucleosynthesis in helium
shells, and on the production of $^{138}$La and $^{180}$Ta by the
$\nu$ process.

\subsection{Effects on Rates of $\nu_e$ and $\bar\nu_e$ Absorption by
Free Nucleons and the $\nu p$ Process}\label{sec-nup}
A $\nu p$ process occurs to produce heavy nuclei during expansion of
the mass elements shown in Fig.~\ref{Fig-tracer} for our supernova model
\cite{Huther:2013dza}. This process requires that matter has $Y_e>0.5$ 
and that significant $\bar\nu_e$ absorption by protons occurs when matter 
evolves through the temperature range of $3\gtrsim T\gtrsim 1$~GK.
Consequently, an important input is the rates of $\nu_e$ and $\bar\nu_e$ 
absorption by free nucleons [reactions~(\ref{eq-nuen}) and (\ref{eq-nuep})].
In the absence of neutrino oscillations, these rates in a mass element
reaching radius $r_m$ at time $t$ can be calculated as
\begin{subequations}
\begin{align}
&\lambda^0_{\nu N}(r_m,t)=\frac{1}{(2\pi)^2}\int E^2f_\nu(t,E,u,r_m)
\sigma_{\nu N}(E)dEdu\\
&=\frac{R_d^2}{8\pi^2r_m^2}\int E^2f_\nu(t,E,u_d,R_d)
\sigma_{\nu N}(E)dEdv/u,\label{eq-rnu}
\end{align}
\end{subequations}
where $\sigma_{\nu N}$ stands for the absorption cross section
$\sigma_{\nu_e n}$ or $\sigma_{\bar\nu_e p}$ given by
\begin{subequations}\label{eq-crosection}
\begin{align}
\sigma_{\nu_en}(E)&=\sigma_0\left(\frac{E+\Delta}{\rm{MeV}}\right)^2
\left[1+1.055\times 10^{-3}\left(\frac{E}{\rm{MeV}}\right)\right], \\
\sigma_{{\bar \nu}_ep}(E)&=\sigma_0\left(\frac{E-\Delta}{\rm{MeV}}\right)^2
\left[1-7.669\times 10^{-3}\left(\frac{E}{\rm{MeV}}\right)\right].
\end{align}
\end{subequations}
In the above equations, $\sigma_0=0.934\times 10^{-43}$~cm$^2$ and
$\Delta=m_n-m_p=1.293$~MeV is the neutron-proton mass difference. 
These cross sections take into account weak magnetism and nucleon recoil 
but neglect the electron rest mass $m_e$ in comparison with $E$ 
\cite{Horowitz:1999fe}.

\begin{figure*}
\includegraphics*[width=1.3\columnwidth, angle=-90]{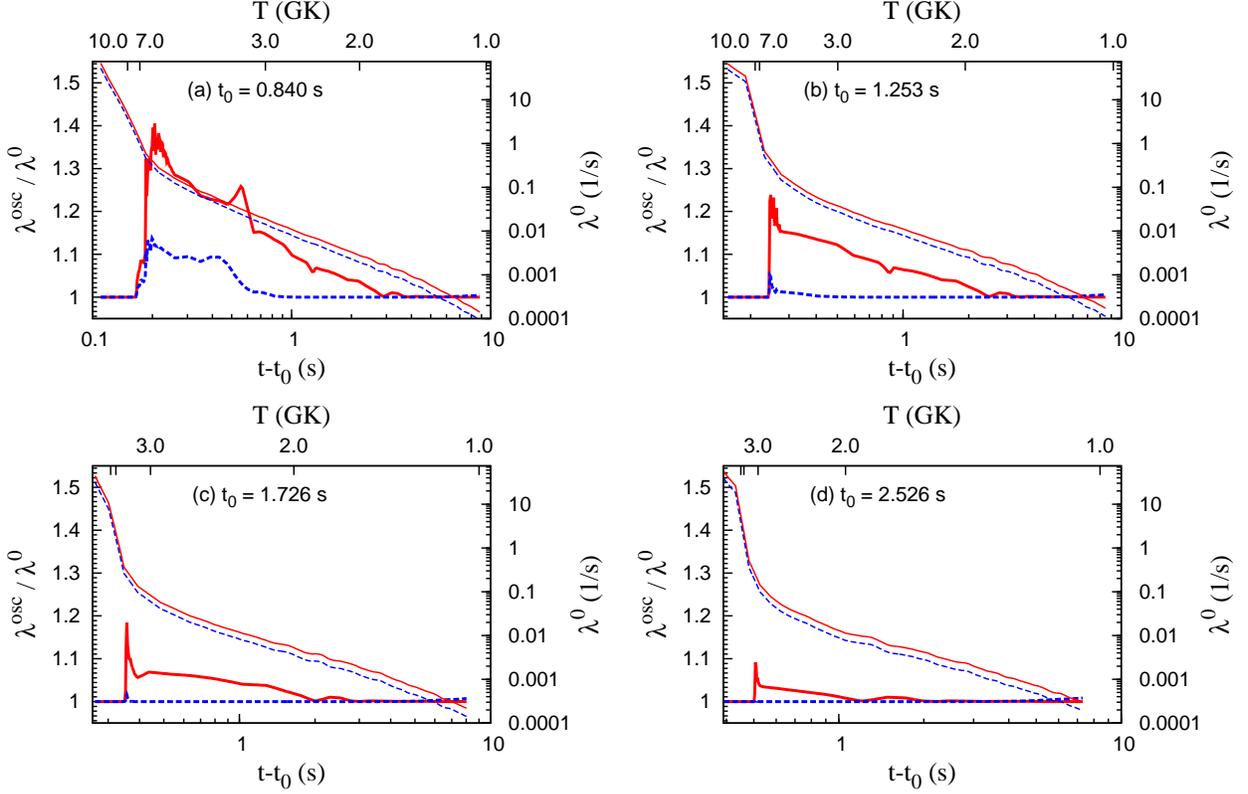}
\caption{Rates $\lambda^0_{\nu N}$ of $\nu_e$ (thin solid red curve) and 
$\bar\nu_e$ (thin dashed blue curve) absorption on free nucleons 
in the absence of neutrino oscillations as functions
of time $t-t_0$ during expansion of four mass elements ejected from the
proto-neutron star at $t_0=0.840$ (a), 1.253 (b), 1.726 (c), and 2.526~s (d),
respectively (both $t$ and $t_0$ are in terms of $t_{\rm pb}$). The 
corresponding thick curves show the ratios 
$\lambda_{\nu N}^{\rm osc}/\lambda^0_{\nu N}$ of absorption rates with and 
without neutrino oscillations. The times at which the temperature of a mass 
element reaches $T=10$, 7, 3, 2, and 1~GK, respectively, are also indicated.
\label{Fig-rate}}
\end{figure*}

The $\nu p$ process occurs at $r_m<10^4$~km (see Fig.~\ref{Fig-tracer}a).
The time for neutrinos to travel from $r=R_d$ to a mass element at these
radii is much shorter than the evolution timescale for the angular 
and energy distributions of neutrinos at emission. Therefore, we have used
$f_\nu(t,E,u,r_m)=f_\nu(t,E,u_d,R_d)$ in Eq.~(\ref{eq-rnu}). The neutrino
travel time is also much shorter than the evolution timescale for the profile
of $n_e(r)$. Consequently, the results on neutrino flavor evolution we have
calculated for each time snapshot of our supernova model can be directly
applied to obtain the rates of $\nu_e$ and $\bar\nu_e$ absorption by 
free nucleons in the presence of neutrino oscillations.
These rates can be calculated as, e.g.,
\begin{align}\label{eq-rate_col}
\lambda_{\nu_en}^{\rm osc}(r_m,t)&=\frac{R_d^2}{8\pi^2r_m^2}\sum_{\alpha}
\int E^2f_{\nu_\alpha}(t,E,u_d,R_d)\sigma_{\nu_en}(E)\nonumber\\
&\times P_{\nu_\alpha\nu_e}(t,E,v,r_m)dEdv/u.
\end{align}

In Fig.~\ref{Fig-rate}, we show $\lambda^0_{\nu N}(r_m,t)$ and 
$\lambda_{\nu N}^{\rm osc}(r_m,t)/\lambda^0_{\nu N}(r_m,t)$ as functions of
$t-t_0$ for the mass elements that are ejected from the proto-neutron star
at $t_0=0.840$, 1.253, 1.726, and 2.526~s, respectively (both $t$ and 
$t_0$ are in terms of $t_{\rm pb}$). It can be seen that 
the overall effect of neutrino oscillations is to enhance both $\nu_e$ and 
$\bar\nu_e$ absorption rates. As shown in Fig.~\ref{Fig-swap_r500}, 
flavor conversion mostly takes place between $\nu_e$ ($\bar\nu_e$) and  
$\nu_y$ ($\bar\nu_y$) with relatively high energies of $E>10$~MeV.
Because the average $\nu_y$ ($\bar\nu_y$) energy is higher, there are more
high-energy $\nu_y$ ($\bar\nu_y$) than $\nu_e$ ($\bar\nu_e$) and the
net effect of flavor conversion is to increase the $\nu_e$ ($\bar\nu_e$) 
absorption rate. In addition, because flavor conversion mostly occurs 
in the neutrino sector (see Fig.~\ref{Fig-swap_r500}), the increase in the 
$\nu_e$ absorption rate is larger than that in the $\bar\nu_e$ absorption rate.
In the region relevant for the $\nu p$ process, the effect of flavor conversion 
on the $\nu_e$ ($\bar\nu_e$) absorption rate also 
diminishes with time and essentially stops at $t_{\rm pb}>5$~s (1.5~s) as
can be seen from Fig.~\ref{Fig-swap_r500}. 
Consequently, the increase in the $\nu_e$ 
($\bar\nu_e$) absorption rate due to neutrino oscillations is negligible for the 
mass elements ejected at $t_0>2.526$ (1.253)~s.

We indicate the times at which the temperature of a mass element reaches
$T=10$, 7, 3, 2, and 1~GK, respectively, in Fig.~\ref{Fig-rate}. To affect the 
setting of $Y_e$ for a mass element, neutrino oscillations must occur when 
its temperature is $T\gtrsim 7$~GK, for which free nucleons dominate its
composition. However, neutrino oscillations start to affect the $\nu_e$ and
$\bar\nu_e$ absorption rates at $T\lesssim 7$~GK for all the mass elements
shown in Fig.~\ref{Fig-rate}, and therefore, have little impact on the setting
of their $Y_e$. The significant increase in the rate of $\nu_e$ absorption by
neutrons at $T\lesssim 7$~GK does not affect nucleosynthesis because
the neutron abundance drops rapidly with decreasing temperature.
On the other hand, when the mass elements evolve through
the temperature range of $3\gtrsim T\gtrsim 1$~GK and $\bar\nu_e$ absorption
by protons is instrumental to the ongoing $\nu p$ process \cite{Frohlich:2005ys},
neutrino oscillations have essentially no effect on the $\bar\nu_e$ absorption rate
(see Fig.~\ref{Fig-rate}). Therefore, the $\nu p$ process in our supernova model
is not affected by neutrino flavor evolution including collective oscillations.

\subsection{Effects on Rates of $\nu_e$ and $\bar\nu_e$ Absorption by 
$^4$He and Neutrino-Induced Nucleosynthesis}\label{sec-nuinduce}

\begin{figure}
\includegraphics*[width=0.6\columnwidth, angle=-90]{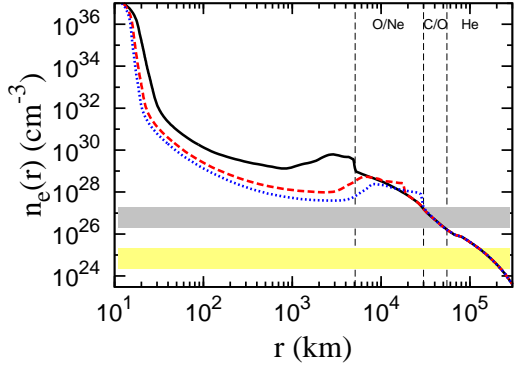}
\caption{Snapshots of the electron number density $n_e$ as a function of 
radius for $t_{\rm pb}=1.0$ (black solid curve), 3.0 
(red dashed curve), and 5.0~s (blue dotted curve), respectively. 
The gray and yellow bands indicate the range of $n_{e,\rm MSW}^{13}$ 
and $n_{e,\rm MSW}^{12}$ for $5<E<100$~MeV, respectively. 
The positions of the O/Ne, C/O, and He layers are
marked by the vertical lines. The bump in $n_e$ at 
$r\sim 10^3$--$10^4$~km corresponds to matter that has been 
shocked recently. \label{Fig-MSW}}
\end{figure}

The neutrino-induced nucleosynthesis occurs typically in 
the C/O and He layers of supernovae located at $r\sim 10^5$~km
for massive stars (see Fig.~\ref{Fig-MSW}). In those layers, the electron 
number density $n_e(r)$ is comparable to or less than the 
MSW resonant density 
$n_{e,\rm MSW}^{\rm (ij)}(E)\equiv(|\Delta m_{\rm ji}^2|\cos{2\theta_{ij}})/(2\sqrt{2} E G_F)$
for relevant neutrino energies of $\sim 5$--100~MeV.
Thus, the detailed neutrino flavor evolution including the MSW effect has
to be considered. In our 18~$M_\odot$ supernova model, the shock reaches 
the MSW resonance region and the C/O layer at 
$t_{\rm pb}\approx 5$~s as shown in Fig.~\ref{Fig-MSW}.
We have checked that for $t_{\rm pb}<5$~s,
C/O and He layers remain static and 
the neutrino flavor transformation through the MSW effect is adiabatic,
i.e., $(\Delta m^2_{\rm ji}\sin^2{2\theta_{\rm ij}})/(2E\cos{2\theta_{\rm ij}})
/|d(\ln{n_e})/dr|_{\rm res} \gg 1$.
The neutrino energy spectra at
$n_e(r)\lesssim n_{e,\rm MSW}^{\rm (13)}(E)$ can then 
be calculated as follows.

At a radius outside the region of collective oscillations but before
the MSW resonances [$n_e(r)\gg n_{e,\rm MSW}^{\rm 13}(E)$], the 
flavor conversion probability can be approximated by
$P_{\nu_\beta\nu_\alpha}(t,E,u_d,r)\approx
P_{\nu_\beta\nu_\alpha}(t,E,u_d,r_f)$. Here $r_f$ is the radius where
collective neutrino oscillations have ceased. Practically, we take 
$r_f=500$~km. Furthermore, for $r\gg R_d$, we can use the
approximation $u\approx 1$, for which the neutrino interaction rates 
including the effects of collective oscillations depend on radius only through 
the geometrical factor $R_d^2/r^2$ [see e.g., Eq.~(\ref{eq-rate_col})].
Thus, it is useful to define the angle-integrated neutrino energy spectra,
\begin{equation}
f^{\rm (i)}_{\nu_\alpha}(t,E)\equiv
\sum_{\beta}\int
f_{\nu_\beta}(t,E,u_d,R_d)P_{\nu_\beta \nu_\alpha}(t,E,u_d,r_f)u_d du_d,
\end{equation}
as the radius-independent spectra before neutrinos enter the MSW region
of flavor evolution. As these spectra correspond to the region 
of $n_e(r)\gg n_{e,\rm MSW}^{\rm 13}(E)$, where 
the effective Hamiltonian is nearly diagonal in the flavor
basis, the flavor eigenstates $|\nu_\alpha\rangle$ and the in-medium mass 
eigenstates $|\nu_i^m \rangle$ that diagonalize the effective Hamiltonian 
can be approximately related by
$|\nu_i^m \rangle = \sum_{\alpha}R_{i \alpha} |\nu_\alpha\rangle$. 
For neutrinos, the only non-zero components of $R_{i \alpha}$ are
$R_{3e}=R_{1x}=R_{2y}=1$ for the normal mass hierarchy (NH)
and $R_{2e}=R_{1x}=R_{3y}=1$ for the inverted mass hierarchy (IH). 
For antineutrinos,
$R_{1e}=R_{2x}=R_{3y}=1$ ($R_{3e}=R_{2x}=R_{1y}=1$)
are the only non-zero components for the NH (IH).

\begin{figure}
\centering
\includegraphics*[width=0.7\columnwidth, angle=-90]{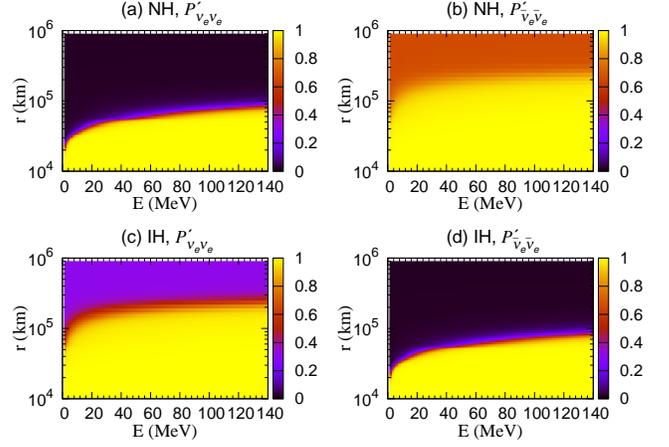}
\caption{Effective probabilities $P^\prime_{\nu_e\nu_e}$ and 
$P^\prime_{\bar\nu_e\bar\nu_e}$ following adiabatic MSW flavor conversion
as functions of $r$ and $E$ for both the NH and IH.}
\label{Fig-peemsw}
\end{figure} 

\begin{figure}
\centering
\includegraphics*[width=0.45\columnwidth, angle=-90]{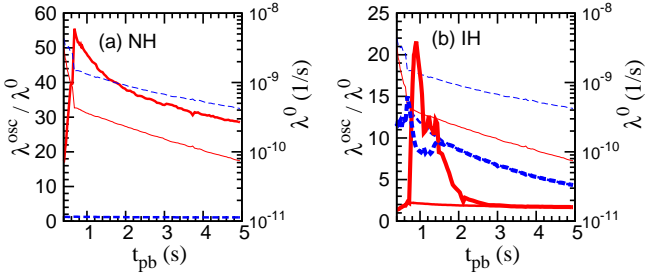}
\caption{(a) The ratio $\lambda^{\rm osc}/\lambda^0$ (thick curves) and 
the rate $\lambda^0$ without neutrino oscillations (thin curves) for 
charged-current $\nu_e$ (red solid curves) and $\bar\nu_e$ (blue dashed curves)
interactions on $^4$He at $r=10^5$~km as functions of $t_{\rm pb}$ for the NH.
(b) Same as (a) but for the IH. The thickest curves are for the full case including
both collective neutrino oscillations and MSW flavor transformation while the
thicker curves are for the case of pure MSW flavor transformation.}
\label{Fig-rate_t}
\end{figure}

For adiabatic MSW flavor evolution, neutrinos initially
in an in-medium mass eigenstate remain in the corresponding mass eigenstate
at later times. As a result, at time $t$ and radius $r$, the neutrino energy spectra 
including the effects of adiabatic MSW flavor transformation can be written as
\begin{align}
{\tilde f}_{\nu_\alpha}(t,E,r) & =
\sum_k\sum_{\beta}{\mathsf P}_{\alpha k}(E,r)R_{k\beta}f^{\rm (i)}_{\nu_\beta}(t^\prime,E) \nonumber \\ 
& \equiv \sum_\beta P^\prime_{\nu_\alpha\nu_\beta}(E,r)f^{\rm (i)}_{\nu_\beta}(t^\prime,E),
\end{align}
where ${\mathsf P}_{\alpha k}(E,r)=|\langle\nu_\alpha|\nu_k^m(E,r)\rangle|^2$ is
the probability that the in-medium mass eigenstate $|\nu_k^m(E,r)\rangle$ at radius $r$
coincides with the flavor eigenstate $|\nu_\alpha\rangle$, and
$P^\prime_{\nu_\alpha\nu_\beta}(E,r)=\sum_k{\mathsf P}_{\alpha k}(E,r)R_{k\beta}$ 
is the effective probability to convert a $\nu_\beta$ at time $t^\prime\approx t-r/c$ 
into a $\nu_\alpha$ at time $t$ and radius $r$ by adiabatic MSW flavor transformation.
Note that ${\mathsf P}_{\alpha k}$ is independent of
time for $t_{\rm pb}<5.0$~s before the shock arrives in the MSW region of flavor evolution. 
We numerically derive the in-medium mass eigenstates $|\nu_k^m(E,r)\rangle$ by
diagonalizing $H_{\rm v}(E)+H_e(r)$ with $n_e(r)$ from the 
supernova progenitor model \cite{Woosley:2002zz}.
We show $P^\prime_{\nu_e\nu_e}$ and $P^\prime_{\bar\nu_e\bar\nu_e}$ as functions
of $r$ and $E$ in Fig.~\ref{Fig-peemsw}.

The rate of neutrino interaction on a target nucleus $A$ at radius $r$ 
can then be calculated as
\begin{equation}\label{eq-rate_MSW}
\lambda_{\nu_\alpha A}^{\rm osc}(r,t)=\frac{R_d^2}{4\pi^2r^2}\int E^2 
{\tilde f}_{\nu_\alpha}(t^\prime,E,r)\sigma_{\nu_\alpha A}(E)dE.
\end{equation}
In Fig.~\ref{Fig-rate_t}, we show the rate $\lambda_{\nu_\alpha A}^0$ without
neutrino oscillations and $\lambda_{\nu_\alpha A}^{\rm osc}/\lambda_{\nu_\alpha A}^0$ 
for charged-current $\nu_e$ and $\bar\nu_e$ interactions on $^4$He at $r=10^5$~km. 
The cross sections are fitted to the form $\sigma_\alpha[(E-E_0)/{\rm MeV}]^n$ by using
the spectrally-averaged cross sections listed in Table II of \cite{Gazit:2007jt}. 
We find that $\sigma_\alpha=3.32\times 10^{-48}$~cm$^2$/nucleon, $E_0=19.8$~MeV, 
$n = 4.01$ for $\nu_e+{^4{\rm He}}\rightarrow {^3{\rm He}}+p+e^-$ and 
$\sigma_\alpha=8.19\times 10^{-48}$~cm$^2$/nucleon, $E_0=21.6$~MeV, 
$n = 3.76$ for $\bar\nu_e+{^4{\rm He}}\rightarrow {^3{\rm H}}+n+e^+$.
As collective oscillations are suppressed in our model for the NH, 
flavor transformation in this case occurs purely through the MSW effect, which
only enhances the $\nu_e$ interaction rate (see Fig.~\ref{Fig-peemsw}). 
The time evolution of $\lambda_{\nu_e A}^{\rm osc}/\lambda_{\nu_e A}^0$ basically 
follows the relative change of the high-energy tail of the $\nu_e$ and $\nu_y$ spectra. 
In the case of IH, there is large enhancement in the $\nu_e$ interaction rate for 
$0.7\lesssim t_{\rm pb}\lesssim 2.5$~s and in the $\bar\nu_e$ interaction rate for
$0.7\lesssim t_{\rm pb}\lesssim 5$~s. For $\nu_e$, the enhancement is
due to partial flavor conversion between $\nu_e$ and $\nu_y$ through collective 
oscillations (see Fig.~\ref{Fig-swap_r500}). In contrast, the enhancement for 
$\bar\nu_e$ is mostly due to MSW flavor conversion. In fact, partial flavor conversion 
through collective oscillations results in a slight reduction of the enhancement in the 
$\bar\nu_e$ interaction rate for $0.7\lesssim t_{\rm pb}\lesssim 1.5$~s when compared 
to the case including MSW flavor conversion only.

\begin{figure}
\centering
\includegraphics*[width=0.45\columnwidth, angle=-90]{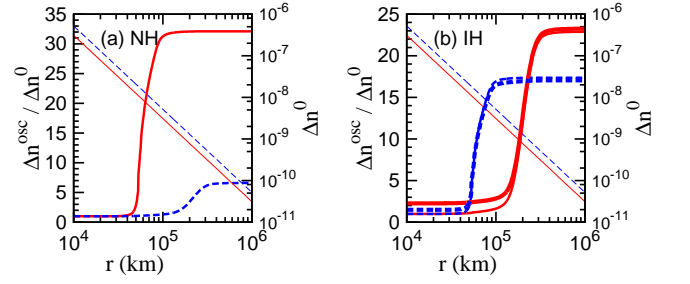}
\caption{(a) The ratio $\Delta n^{\rm osc}/\Delta n^0$ (thick curves) 
and the time-integrated rate $\Delta n^0$ without neutrino oscillations (thin curves)
as functions of $r$ for charged-current $\nu_e$ (red solid curves) 
and $\bar\nu_e$ (blue dashed curves) interactions on $^4$He for the NH.
(b) Same as (a) but for the IH. The thickest curves are for the full case including
both collective neutrino oscillations and MSW flavor transformation while the
thicker curves are for the case of pure MSW flavor transformation.}
\label{Fig-rate_r}
\end{figure}

Neutrino-induced nucleosynthesis in He shells is sensitive to
the time integrated rate of (anti)neutrino absorption on $^4$He, 
$\Delta n(r)\equiv\int \lambda(r,t)dt $. We calculate the
integral for the first 5~s post bounce and show the enhancement factor 
$\Delta n^{\rm osc}(r)/\Delta n^0(r)$ as a function of radius in Fig.~\ref{Fig-rate_r}. 
For the He layer located at $5.5\times 10^4\lesssim r\lesssim 4\times 10^5$~km 
in our model, $\Delta n(r)$ for $\nu_e$ ($\bar{\nu}_e$) absorption on $^4$He
is enhanced by a factor of $\sim 32$ ($\sim 17$) for the NH (IH)
following adiabatic MSW flavor transformation through the 1-3 resonance. 
Before this MSW resonance, collective oscillations for the IH result in an enhancement 
factor of $\sim 2.2$ and $\sim 1.5$ for $\nu_e$ and $\bar\nu_e$ absorption, respectively.
Both the collective and MSW flavor transformations will increase the production 
of $^7$Li and $^{11}$B. To quantify the increase will require a full calculation of
neutrino-induced nucleosynthesis, which is beyond the scope of the present work. 
The impact of MSW flavor transformation on the production of $^7$Li and $^{11}$B
was explored in \cite{Yoshida:2006sk}. However, this previous study assumed harder 
neutrino spectra than calculated by our supernova simulations, which makes it difficult 
to extrapolate their conclusions to our model.

\subsection{Effects on Rates of $\nu_e$ Absorption by 
$^{138}$Ba and $^{180}$Hf and the production of 
$^{138}$La and $^{180}$Ta}\label{sec-nuLaTa}
The rare isotopes $^{138}$La and $^{180}$Ta can be produced 
predominantly by the charged-current interaction of $\nu_e$ on the preexisting 
$^{138}$Ba and $^{180}$Hf, respectively.
The main production region is in the O/Ne shell of our supernova progenitor
and the yields sensitively depend on the neutrino ``temperature'' \cite{Heger:2003mm}.
During the first 5~s post bounce, $n_e(r)$ in the O/Ne shell is much larger than
$n_{e,\rm MSW}^{\rm 13}(E)$ for typical neutrino energies, and
the neutrino interaction rates on nuclei can be
calculated similarly to Eq.~(\ref{eq-rate_MSW}) but with 
${\tilde f}_{\nu_\alpha}(t^\prime,E,r)$ replaced by $f^{\rm (i)}_{\nu_\alpha}(t,E)$.
In this case, flavor conversion results from collective oscillations only and
the ratio $\lambda^{\rm osc}_{\nu_\alpha A}/\lambda^0_{\nu_\alpha A}$ is
independent of radius due to cancellation of the $1/r^2$ dependence for
each rate.

\begin{figure}
\centering
\includegraphics*[width=0.45\columnwidth, angle=-90]{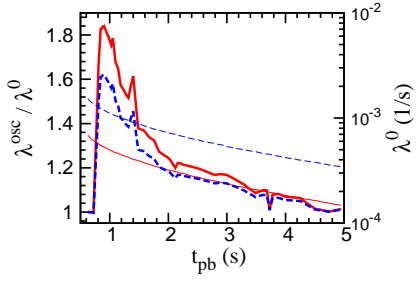}
\caption{The ratios $\lambda^{\rm osc}/\lambda^0$ for charged-current interactions of 
$\nu_e$ on $^{138}$Ba (red solid curves) and $^{180}$Hf (blue dashed curves) 
as functions of $t_{\rm pb}$ for the IH. The rates $\lambda^0$ at 
$r=10^4$~km without oscillations (thin curves) are also shown.}
\label{Fig-rate_LaTa}
\end{figure}

We have calculated the rates of reactions~(\ref{eq-nuBa}) and (\ref{eq-nuHf})
with and without neutrino oscillations, using cross sections with power-law 
dependence on neutrino energy fitted to the spectrally-averaged cross sections 
adopted in \cite{Heger:2003mm}. The enhancement factors
$\lambda^{\rm osc}_{\nu_\alpha A}/\lambda^0_{\nu_\alpha A}$ for
these rates for the IH are shown in Fig.~\ref{Fig-rate_LaTa}. Although collective oscillations
enhance the rates of $\nu_e$ interactions on $^{138}$Ba and $^{180}$Hf
by up to $\sim 80\%$ and $\sim 60\%$, respectively, during $0.8<t_{\rm pb}<5$~s, 
the time-integrated rates for the first 5~s post bounce are increased by only 
$\sim 11.5$\% and $\sim 8.5$\%, respectively. This is because the neutrino luminosity 
is much higher but collective oscillations are suppressed during the first 0.8~s post bounce.
Consequently, while $\nu_e$ interactions during the cooling phase are substantially enhanced 
by collective oscillations for the IH, they still contribute sub-dominantly to the total production of 
$^{138}$La and $^{180}$Ta.

\section{Effects of Flavor Oscillations on Neutrino Signals}\label{sec-signal}
For predicting neutrino signals from a Galactic supernova described by our model, 
the event rate for a particular neutrino interaction with target particles $i$ in a detector 
can be approximately calculated as
\begin{equation}
R_i(t)=\frac{N_iR_d^2}{4\pi^2d^2}\sum_\alpha\int_{E_{\rm th}}^{\infty} 
E^2f^{\rm (f)}_{\nu_\alpha}(E,t)\sigma_{\nu_\alpha i}(E)dE,
\end{equation}
where $N_i$ is the total number of target particles $i$ in the detector, $d$
is the distance from the supernova to the Earth, and $E_{\rm th}$ is 
the threshold energy for the interaction.
Here we have assumed 100\% detection efficiency for simplicity. 
For the neutrino energy spectra $f_{\nu_\alpha}^{\rm (f)}$
at the Earth, we neglect the slight modification by the Earth-matter effect 
\cite{Borriello:2012zc} and assume
$f_{\nu_\alpha}^{\rm (f)}=\tilde{f}_{\nu_\alpha}({n_e=0})$.
Specifically, we take
\begin{subequations}{\label{eq-numswnh}}
\begin{align}
f^{(\rm f)}_{\nu_e}& =s^2_{13}f_{\nu_e}^{(\rm i)}+c^2_{12}c^2_{13}f_{\nu_x}^{(\rm i)}
+s^2_{12}c^2_{13}f_{\nu_y}^{(\rm i)},\\
f^{(\rm f)}_{\nu_x}&=s^2_{12}f_{\nu_x}^{(\rm i)}+c^2_{12}f_{\nu_y}^{(\rm i)},\\
f^{(\rm f)}_{\nu_y}&=c^2_{13}f_{\nu_e}^{(\rm i)}+c^2_{12}s^2_{13}f_{\nu_x}^{(\rm i)}
+s^2_{12}s^2_{13}f_{\nu_y}^{(\rm i)},\\
f^{(\rm f)}_{\bar\nu_e}&=c^2_{12}c^2_{13}f_{\bar\nu_e}^{(\rm i)}+s^2_{12}c^2_{13}f_{\bar\nu_x}^{(\rm i)}
+s^2_{13}f_{\bar\nu_y}^{(\rm i)},\\
f^{(\rm f)}_{\bar\nu_x}&=s^2_{12}f_{\bar\nu_e}^{(\rm i)}+c^2_{12}f_{\bar\nu_x}^{(\rm i)},\\
f^{(\rm f)}_{\bar\nu_y}&=c^2_{12}s^2_{13}f_{\bar\nu_e}^{(\rm i)}+s^2_{12}s^2_{13}f_{\bar\nu_x}^{(\rm i)}
+c^2_{13}f_{\bar\nu_y}^{(\rm i)},
\end{align}
\end{subequations}
for the NH, and
\begin{subequations}{\label{eq-numswih}}
\begin{align}
f^{(\rm f)}_{\nu_e}&=s^2_{12}c^2_{13}f_{\nu_e}^{(\rm i)}+c^2_{12}c^2_{13}f_{\nu_x}^{(\rm i)}
+s^2_{13}f_{\nu_y}^{(\rm i)},\\
f^{(\rm f)}_{\nu_x}&=c^2_{12}f_{\nu_e}^{(\rm i)}+s^2_{12}f_{\nu_x}^{(\rm i)},\\
f^{(\rm f)}_{\nu_y}&=s^2_{12}s^2_{13}f_{\nu_e}^{(\rm i)}+c^2_{12}s^2_{13}f_{\nu_x}^{(\rm i)}
+c^2_{13}f_{\nu_y}^{(\rm i)},\\
f^{(\rm f)}_{\bar\nu_e}&=s^2_{13}f_{\bar\nu_e}^{(\rm i)}+s^2_{12}c^2_{13}f_{\bar\nu_x}^{(\rm i)}
+c^2_{12}c^2_{13}f_{\bar\nu_y}^{(\rm i)},\\
f^{(\rm f)}_{\bar\nu_x}&=c^2_{12}f_{\bar\nu_x}^{(\rm i)}+s^2_{12}f_{\bar\nu_y}^{(\rm i)},\\
f^{(\rm f)}_{\bar\nu_y}&=c^2_{13}f_{\bar\nu_e}^{(\rm i)}+s^2_{12}s^2_{13}f_{\bar\nu_x}^{(\rm i)}
+c^2_{12}s^2_{13}f_{\bar\nu_y}^{(\rm i)},
\end{align}
\end{subequations}
for the IH. In the above equations, $c_{ij}$ and $s_{ij}$ stand for $\cos{\theta_{ij}}$ and 
$\sin{\theta_{ij}}$, respectively.

\begin{figure}
\centering
\includegraphics*[width=2.0\columnwidth, angle=-90]{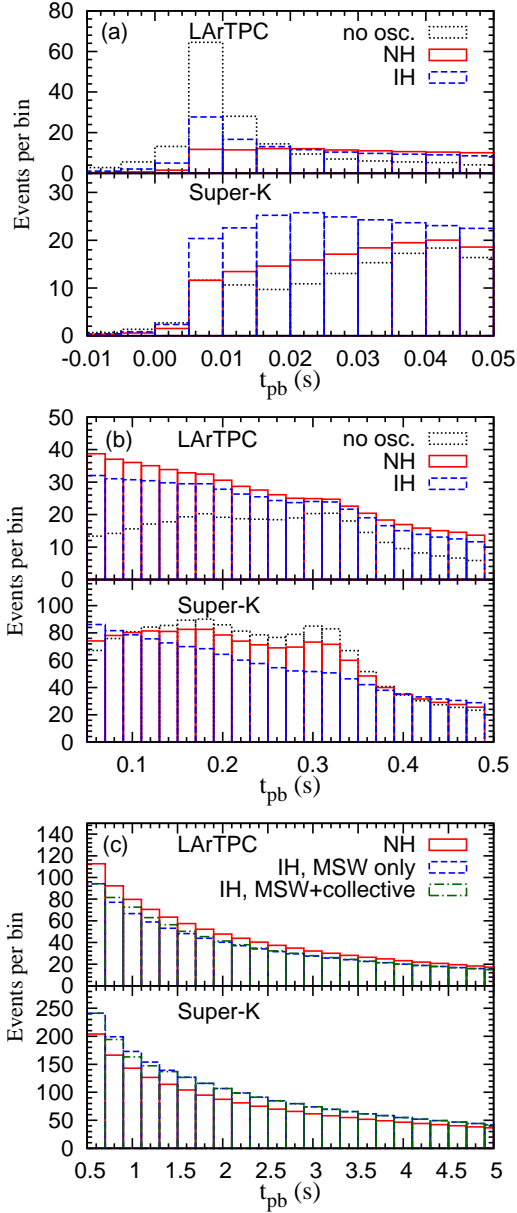}
\caption{The number of all neutrino events per time bin as a function of time
for an LArTPC detector and the Super-K detector, respectively, during 
(a) the neutronization burst, (b) the accretion phase, and (c) the proto-neutron 
star cooling phase of a Galactic supernova at a distance of 10~kpc.}
\label{fig-sigtb}
\end{figure}

We have calculated the expected neutrino signals in a 34~kton
liquid argon time projection chamber (LArTPC) detector and the Super-Kamiokande (Super-K)
detector for a Galactic supernova at $d=10$~kpc. 
The included interaction channels are
\begin{subequations}
\begin{align}\label{eq-nueAr}
\nu_e + {^{40}{\rm Ar}} & \rightarrow e^- + {^{40}{\rm K^*}},\\
\bar\nu_e + {^{40}{\rm Ar}} & \rightarrow e^+ + {^{40}{\rm Cl^*}},\\
\nu_\alpha + e^- & \rightarrow \nu_\alpha + e^-,
\end{align}
\end{subequations}
for the LArTPC detector, and
\begin{subequations}
\begin{align}\label{eq-nuebp}
\bar\nu_e + p & \rightarrow e^+ + n,\\
\nu_\alpha + e^- & \rightarrow \nu_\alpha + e^-,\\
\nu_\alpha + {^{16}{\rm O}} & \rightarrow \nu_\alpha + {^{16}{\rm O}},\\
\nu_e + {^{16}{\rm O}} & \rightarrow e^- + {^{16}{\rm F^*}},\\
\bar\nu_e + {^{16}{\rm O}} & \rightarrow e^+ + {^{16}{\rm N^*}},
\end{align}
\end{subequations}
for the Super-K detector.
The cross sections for neutrino interactions on $^{16}$O and $^{40}$Ar have been
computed in \citep{Kolbe:2003ys}. The numerical values for all cross
sections are taken from the data compiled in \cite{Scholberg:2012id}.
In Fig.~\ref{fig-sigtb}, we show the number of all neutrino events in time bins of 
5, 20, and 200~ms for the neutronization burst, the accretion phase, and the 
proto-neutron star cooling phase, respectively, as a function of time. 

The neutronization burst could be readily seen in the LArTPC detector if there 
were no neutrino oscillations. However, the burst completely disappears
for the NH. For the IH, there might still be a chance to identify the burst from the 
dominant interaction channel of $\nu_e$ capture on $^{40}$Ar. This can be understood
by examining Eqs.~(\ref{eq-numswnh}) and (\ref{eq-numswih}).
With $c^2_{12}c^2_{13}\sim 0.69$, $s^2_{12}c^2_{13}\sim 0.31$, 
and $s^2_{13}\sim 0$, we have $f^{(f)}_{\nu_e}\sim f_{\nu_{\mu(\tau)}}$ for the NH
and $f_{\nu_e}^{(f)}\sim 0.31 f_{\nu_e}+0.69 f_{\nu_{\mu(\tau)}}$ for the IH
when there are no collective neutrino oscillations. As the $\nu_{\mu(\tau)}$ flux is much 
smaller compared to $\nu_e$ during the neutronization burst,
the number of events with oscillations is strongly limited.
For the Super-K detector, the dominant detection channel is $\bar\nu_e$ absorption on 
protons and the $\bar\nu_e$ flux at the detector is 
$f_{\bar\nu_e}^{(f)}\sim 0.69 f_{\bar\nu_e}+0.31 f_{\bar\nu_{\mu(\tau)}}$ 
for the NH and $f_{\bar\nu_e}^{(f)}\sim f_{\bar\nu_{\mu(\tau)}}$ for the IH. Thus,
there will be more events for the IH because $\bar\nu_{\mu(\tau)}$ have a higher 
average energy than $\bar\nu_e$. Note that if the distance to the supernova is much 
closer than 10~kpc, there will be many more events during the neutronization burst
so that the rising time of the $\bar\nu_e$ signal may be used to distinguish the 
neutrino mass hierarchy \cite{Serpico:2011ir}.

During the accretion phase, the time profiles of
the events in the two detectors behave quite differently for either
mass hierarchy. For the NH, the $\bar\nu_e$ luminosity plateau at emission 
(see Fig.~\ref{Fig-lumin}) can still be observed in the Super-K detector, but in
the LArTPC detector, the event rate basically follows the
decreasing $\nu_{\mu(\tau)}$ luminosity at emission.
For the IH, both detectors will see a decreasing number of 
events per bin with time, but the rate of decrease
will be larger for the Super-K detector. If the time profiles of (anti-)neutrino luminosities 
at emission in our model are generic for supernovae, it will be possible to
utilize the time profiles of the events in the two detectors, e.g., by forming
a ratio of the respective number of events per time bin,
to infer the neutrino mass hierarchy.

During the cooling phase, the time profiles of the events are
rather similar for both mass hierarchies and whether collective oscillations
are included or not. As the 1-2 mixing is entirely suppressed in
the region of collective oscillations, it is straightforward to 
show from Eqs.~(\ref{eq-numswnh}) and (\ref{eq-numswih}) that the neutrino signals 
will always be between the cases of pure adiabatic MSW flavor transformation for the 
NH and IH. The similarity of the event time profiles in different scenarios of neutrino
oscillations is further enhanced by the convergence of
all neutrino spectra during the cooling phase.

\begin{figure}
\centering
\begin{tabular}{c}
\includegraphics*[width=0.85\columnwidth, angle=-90]{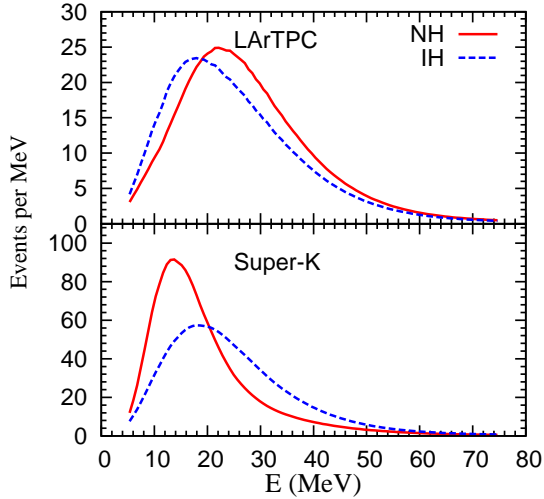} \\
\end{tabular}
\caption{Time-integrated energy spectra of neutrino events during the first 0.5~s
for the LArTPC and Super-K detectors.
\label{fig-sigeacc}}
\end{figure}

From the above discussion, it is clear that the neutrino signals during 
the accretion phase give a better chance to resolve the neutrino mass
hierarchy. During this phase, there are substantial differences in the
(anti-)neutrino spectra between the electron and $\mu(\tau)$ flavors 
and the total number of expected events is also relatively large
compared to the neutronization burst. It is 
thus interesting to examine the energy spectra of the neutrino events
during the accretion phase for both mass hierarchies.
In Fig.~\ref{fig-sigeacc}, we show the time-integrated neutrino spectra 
during the first 0.5 s for the LArTPC and Super-K detectors, 
assuming that the neutrino energy can be inferred from detection. 
It can be seen that for the NH, the spectrum in the Super-K detector
has a much lower average energy and smaller spread compared
with the spectrum in the same detector for the IH or the spectrum in
the LArTPC detector for either mass hierarchy. Thus, assuming that
collective neutrino oscillations are indeed strongly suppressed
during the accretion phase of supernova explosion, we expect that
the NH can be clearly identified if a large difference in the neutrino 
spectra is detected between the Super-K and LArTPC detectors.
This identification can be made more robust by using
the different time profiles for the two detectors in addition.

\section{Conclusions}\label{sec-summary}
We have shown that in order to understand the effects of 
neutrino flavor oscillations on supernova nucleosynthesis and on the 
neutrino signals, a detailed calculation including
both collective oscillations and the subsequent MSW
flavor transformation has to be carried out to derive the flavor 
conversion probabilities as functions of
the emission time, energy, propagation angle, and arrival 
radius for a neutrino. We have explicitly described a detailed scheme
of calculation and implemented it by employing the time-dependent 
neutrino spectra and electron density profiles 
from a spherically-symmetric supernova model with an $18\,M_\odot$ progenitor. 
We find that collective neutrino oscillations are not only sensitive to
the detailed neutrino energy and angular distributions at emission, but
also to the time evolution of these distributions and the electron density profile
due to the non-linear nature of the flavor evolution equation as
discussed in Sec.~\ref{sec-nuosc}.

We have shown that in our supernova model, 
collective neutrino oscillations happen only for the IH, mostly in the neutrino
sector during the first 5~s of the proto-neutron star cooling phase, 
due to the dominant emission of $\bar\nu_{\mu(\tau)}$ over $\bar\nu_e$.
The radius/temperature at which these oscillations occur is in
general too far/low to have a direct impact on $Y_e$ before free nucleons
are assembled into $\alpha$ particles.
In addition, there is no effect on the rate of $\bar\nu_e$ capture 
on protons for the temperature range of $1 \lesssim T \lesssim 3$~GK,
which is relevant for the $\nu$p process in the neutrino-driven wind. Thus, 
the outcome of this process in our supernova model is not affected by
collective neutrino oscillations as discussed in Sec.~\ref{sec-nup}.
However, collective oscillations may still enhance somewhat the production of 
the rare isotopes $^{138}$La and $^{180}$Ta as shown in Sec.~\ref{sec-nuLaTa}.
We have also calculated the MSW flavor transformation that happens in the C/O and He shells
of the supernova for the first 5~s of post-bounce time,
in order to evaluate the impact of neutrino oscillations
on the neutrino-induced nucleosynthesis in the He shell and on the neutrino signals. 
We find that the charged-current interaction rates of $\nu_e$ ($\bar\nu_e$)
on $^4$He are greatly enhanced by the MSW flavor transformation for the NH (IH). 
This may impact the production of $^{7}$Li and $^{11}$B as
studied in \cite{Yoshida:2006sk}.

For the neutrino signals from our supernova model, we have calculated the number of
events per time bin for different neutrino emission phases 
for the Super-Kamiokande detector and a hypothetical 34~kton
liquid argon detector. The results suggest that for a Galactic supernova of this kind,
the events from the neutronization burst may not be enough to
identify the burst. However, it may be possible to use the time profiles of events during 
the accretion phase and the associated neutrino spectra in these two detectors to 
infer the yet-unknown neutrino mass hierarchy.
For the cooling phase, we have shown that the effects of collective 
oscillations are in general small, limited by the difference between the cases of 
pure MSW flavor transformation for the NH and IH.

Although our results seem to suggest that collective neutrino
oscillations do not have a large impact on either the nucleosynthesis or the
neutrino signals, cautions must be mentioned
as there are a number of issues in modeling such oscillations
in supernovae. First, in our treatment here, we have
assumed azimuthal symmetry around the radial direction for collective oscillations.
The effect of relaxing this symmetry \cite{Raffelt:2013rqa} needs 
to be further examined and may require a full six-dimensional calculation for 
neutrinos emitted in different parts of the proto-neutron star surface.
Second, it has been suggested recently that the coherence between different neutrino
mass eigenstates may not be maintained by the time collective oscillations
occur, due to the very small wave-packet size of neutrinos at 
production \cite{Akhmedov:2014ssa}. Detailed
examination of this effect in connection with the processes 
of neutrino production is thus required. Third, although 
studies of flavor instabilities suggest that the contribution to the effective Hamiltonian
from those neutrinos scattered by nuclei outside a proto-neutron star, the
so-called neutrino ``halo'', may further suppress collective oscillations during the
accretion phase \cite{Cherry:2012zw,Sarikas:2012vb}, this remains to be confirmed 
by a self-consistent calculation. Fourth, our view of collective oscillations in particular
and neutrino flavor transformation in general might be changed by
the apparently sub-dominant beyond-the-mean-field contribution \cite{Volpe:2013uxl}, 
the neutrino spin coherence \cite{Vlasenko:2014bva},
and the neutrino magnetic moment \cite{deGouvea:2013zp}. 
Last but not the least, as we have demonstrated in this paper, 
the outcome of neutrino flavor oscillations is sensitive to the input from the supernova
model. For example, any change in neutrino spectra from supernova models with 
improved treatment of neutrino interaction with matter in the decoupling region 
\cite{MartinezPinedo:2012rb,Roberts:2012um} will require reexamination of the problem 
of supernova neutrino oscillations.

In summary, this work represents a small step towards the understanding
of supernova neutrino oscillations and their impact.
In view of the importance of supernova nucleosynthesis 
to the production history of various nuclei in our universe, along with 
the high reward of utilizing neutrino signals from a future Galactic supernova 
to learn about supernova physics and neutrino properties,
further studies that take into account all the issues mentioned above,
not for one, but for a large number of models with different progenitors, 
must be carried out in order to fully understand the
effects of neutrino oscillations in supernovae and the associated rich physics.

\begin{acknowledgments}
M.-R.W. is partly supported by the Alexander von Humboldt Foundation.
 Y.-Z.Q. is partly supported by the US DOE (DE-FG02-87ER40328).
L.H. and G.M.P. are partly supported by the Deutsche Forschungsgemeinschaft
through contract SFB 634, the Helmholtz International Center for FAIR
within the framework of the LOEWE program launched by the state of
Hesse and the Helmholtz Association through the Nuclear Astrophysics
Virtual Institute (VH-VI-417).
T.F. acknowledges support from the Narodowe Centrum Nauki (NCN)
within the "Sonata" program under contract No. UMO-2013/11/D/ST2/02645.
M.-R.W. and G.M.P. gratefully thank Andre Sieverding for
helpful discussions. This work was carried out in part using 
computing resources at the University of Minnesota Supercomputing Institute.
\end{acknowledgments}

%

\end{document}